**Electron-Phonon Coupling and Electron-Phonon Scattering in SrVO$_3$**

*Mathieu Mirjolet, Francisco Rivadulla, Premysl Marsik, Vladislav Borisov, Roser Valentí, and Josep Fontcuberta\**

M. Mirjolet, Prof. J. Fontcuberta
Institut de Ciència de Materials de Barcelona (ICMAB-CSIC), Campus UAB, Bellaterra, 08193, Spain
E-Mail: fontcuberta@icmab.cat

Dr. Francisco Rivadulla
CIQUS, Centro de Investigación en Química Biolóxica e Materiais Moleculares, and Departamento de Química-Física, Universidade de Santiago de Compostela, Santiago de Compostela, 15782, Spain.

Dr. P. Marsik
Department of Physics, Faculty of Science and Medicine, University of Fribourg, Fribourg, CH-1700, Switzerland

Dr. V. Borisov
Department of Physics and Astronomy, Uppsala University, Box 516, Uppsala, SE-75120, Sweden

Prof. R. Valentí
Institut für Theoretische Physik, Goethe-Universität Frankfurt am Main; Frankfurt am Main 60438, Germany

Understanding the physics of strongly correlated electronic systems has been a central issue in condensed matter physics for decades. In transition metal oxides, strong correlations characteristic of narrow *d* bands is at the origin of such remarkable properties as the Mott gap opening, enhanced effective mass, and anomalous vibronic coupling, to mention a few. SrVO$_3$, with V$^{4+}$ in a $3d^1$ electronic configuration is the simplest example of a 3D correlated metallic electronic system. Here, we focus on the observation of a (roughly) quadratic temperature dependence of the inverse electron mobility of this seemingly simple system, which is an intriguing property shared by other metallic oxides. The systematic analysis of electronic transport in SrVO$_3$ thin films discloses the limitations of the simplest picture of *e-e* correlations in a Fermi liquid; instead, we show that the quasi-2D topology of the Fermi surface and a strong electron-phonon coupling, contributing to *dress* carriers with a *phonon cloud*, play a pivotal role on the reported electron spectroscopic, optical, thermodynamic and transport data. The picture that emerges is not restricted to SrVO$_3$ but can be shared with other 3*d* and 4*d* metallic oxides.

**1. Introduction**

Strong coulomb interactions characteristic of partially occupied narrow 3*d* bands *renormalize* the properties of charge carriers in *Fermi liquids* (FL), resulting, among other effects, in a large increase of their effective mass $m^*_{ee}$, with respect to the effective band mass $m^*_{band}$. Increasing further the carrier density (*n*) or reducing the conduction band width (*W*), may eventually give rise to an emerging insulating state (Mott transition).[1–4]

This framework has been used to rationalize the properties of correlated electronic systems, and particularly of transition metal oxide (TMO) perovskites ABO$_3$ (B is a transition metal). The robustness of the perovskite scaffolding (a 3D network of octahedrally coordinated BO$_6$ polyhedra) allows multiple cation substitutions at A/B sites that modify the electron filling, and B-O bond distances and angles, achieving in some cases a band-filling and bandwidth-driven Mott transition.[5]

The complexity derived from the interplay of all these parameters is schematized in the diagram below (**Figure 1**), where we take as example a TMO containing one single electron in a 3*d*[1] band, e.g. SrVO$_3$. In a cubic BO$_6$ cage, the single electron of the transition metal M occupies the ($d_{xy}$, $d_{xz}$, $d_{yz}$) orbitals of 3*d*-$t_{2g}$ parentage. The resulting band (of width *W*), being partially occupied (1/6), will host a metallic conductivity (Figure 1, center). Within the FL picture, the carrier effective mass would be *renormalized* to $m^*_{ee}$, which will be larger than $m^*_{\text{band}}$. Modification of the bond topology and charge distribution within the $t_{2g}$ band, may promote the opening of a Mott gap. Vanadium $3d^1$ oxides such as $(Sr,Ca)VO_3$ or $VO_2$ fit in this picture, as recently overviewed by Brahlek et al.[6] According to that, the red-shifted plasma frequency ($\omega^*_p \propto (n/m^*)^{1/2}$) of $(Sr,Ca)VO_3$ can be attributed to a large effective mass $m^*$ arising from *e-e* correlations (Figure 1, bottom). In the same vein, an abrupt metal insulator transition occurs in $VO_2$ upon cooling due to the opening of a Mott gap (Figure 1, top). The properties of these materials are thus described within a purely electronic model including correlation effects (Figure 1, bottom-top).

Although the success of this approach has been tremendous, some properties of metallic oxides cannot be described within this framework. For instance, it has been repeatedly reported that the resistivity ($\rho$) and the inverse carrier mobility ($\mu^{-1}$) of $SrVO_3$ (SVO) follow a nearly $T^2$ temperature dependence.[7–10] In fact, $\mu(T) \approx T^{-2}$ is commonly observed in doped Mott insulators, such as $LaTiO_3$ or $SmTiO_3$, but also in doped band insulators (e.g. $SrTiO_{3-x}$ and $(Gd,La,Nb):SrTiO_3$ as recently reviewed by Stemmer el al.[11]), some high-$T_C$ superconductors (HTS), or oxyselenides.[12] It has been emphasized that although $\mu \approx T^{-2}$ may be consistent with the picture of *e-e* scattering in a FL, the dependence of the amplitude of this scattering term on carrier density sharply contradicts expectations based on the FL description of interacting electron systems.[11,13] Therefore, the whole scenario should be revisited.

In this regard, we suggest that the topology of the Fermi surface (FS) could be an important ingredient to this picture, previously overlooked. Along the Γ-X direction, only ($d_{xy}$, $d_{xz}$) orbitals overlap and therefore the FS consists of a 2D cylinder. A similar situation occurs along the orthogonal directions, and thus the FS is formed by three interpenetrated cylinders (Figure 1, right). The relevance of a quasi-2D FS on the carrier mobility is not minor: in anisotropic metals with quasi-cylindrical FS branches, *e-ph* scattering promotes a distinctive ≈ $T^{-2}$ temperature dependence of the carrier mobility, early described in great detail for Bi.[14] The second feature to consider is that carriers in a $3d^1$ TMO move within an ionic lattice background. This implies that the lattice can be polarized around the moving charge. Carriers, in this scenario (Figure 1, left) are dressed with a lattice distortion, that translates into an effective mass ($m^*_{e-ph}$) related to the coupling of the electrons to the lattice (*e-ph*) coupling. In view of the two above-mentioned aspects, SVO is at the crossroad of *e-e* correlations, *e-ph* strong coupling and low dimensional FS (Figure 1), whose interplay requires renewed attention. An important general question is whether the quadratic temperature dependence of the resistivity is indeed an undisputable fingerprint of *e-e* correlations, or if other scenarios should be envisaged.

Here, we aim at revising some of these issues by reporting on electric transport data (resistivity, magnetoresistance, Hall and Seebeck coefficients) of a large set of SrVO₃ (SVO) epitaxial films deposited on different substrates and with different growth conditions, selected to tune the lattice distortions and, expectedly, the lattice dynamics.

It will be first shown that all films have roughly $\mu(T) \approx T^{-2}$ and $\rho(T) \approx A\,T^2$; these temperature dependences are consistent with earlier findings,[7–10] where $A$ was identified as the coefficient corresponding to the temperature-dependent *e-e* scattering rate ($A_{ee}$) in a FL, i.e. $A \equiv A_{ee}$.[15–17] However, from the analysis of the magnitude of $A$ and its carrier density dependence, we conclude that the temperature dependence of $\mu(T)$, in contrast to earlier views, may not originate from *e-e* scattering in a FL. Instead, we show that $\rho(T)$ can be well described by a polaronic model (Figure 1, left panel), where the relevant phonon energy can be tuned *ad-hoc* by epitaxial strain and subsequent lattice distortion. We also show that *e-ph* scattering in a cylindrical 2D FS (Figure 1, right panel) accounts for the observed $\mu(T) \approx T^{-2}$, the temperature-dependent Seebeck coefficient $S(T)$ and the observed magnetoresistance and its Kohler's scaling. We thus conclude that *e-e* interactions and *e-ph* coupling in $3d^1$ TMOs should be taken on an equal footing to account for the experimental mass renormalization, as found in other oxides,[18,19] and that the 2D character of the FS should be explicitly considered to build a comprehensive view of the carrier transport in these seemingly most simple metallic oxides.

## 2. Results

### 2.1. Temperature dependence of the electrical resistivity

Films were grown on different cubic substrates (LaAlO$_3$ (LAO), NdGaO$_3$ (NGO), LSAT and SrTiO$_3$ (STO)) which have different structural mismatch with SVO. Depending on film thickness ($t$), substrates may impose tensile (STO) or compressive (LAO) epitaxial strain on nominally cubic (bulk) SVO, promoting a tetragonal distortion ($c/a \neq 1$, where $(a, c)$ are the cell parameters) as determined by X-ray diffraction (Supporting Information S1). Growth rate and carrier concentration were adjusted by growth atmosphere. **Figure 2**a,b displays some $\rho(T)$ curves, representative of films on different substrates, all having a $t \approx 70$ nm and grown at different $P(Ar)$. Although the growth pressure has a major effect on $\rho(T)$, all films show a metallic-like resistivity, and the $\rho(300\,\text{K}) \approx 85 - 100\,\mu\Omega\,\text{cm}$ for films grown on LSAT, NGO and LAO is similar to state-of-the-art data.[8–10,20–22] The resistivity of the SVO film on STO is vertically shifted due to the presence of planar defects associated to the large tensile mismatch of SVO with the STO substrate,[23] although its slope remains similar to the other films. In Figure 2c, we show $\rho(T)$ of SVO films of different thicknesses on NGO; the resistivity of the films gradually increases as thickness reduces, as commonly found in SVO.[8–10] The residual resistivity ratio (*RRR*) (Supporting Information S2 is as high as $\approx 12$, among the highest values ever reported for PLD-grown SVO films,[24] in optimally grown films (70 nm; $P(Ar) =$ 0.2-0.3 mbar), and decreases down to $RRR \approx 2$ in films grown on poorly matched substrates and/or too-high or too-low $P(Ar)$. Thus, the *RRR* is used in the following as a label for film quality.[9,22,25,26]

All $\rho(T)$ data in Figure 2a-c display a characteristic non-linearity.[8–10,21,22] Data can be roughly described as $\rho(T) = \rho_0 + A\,T^2$ and can be fitted to extract ($\rho_0$, $A$) (see fits in **Figure 5** below). In **Figure 3**a, we show the extracted $A$ values for all samples as a function of the carrier concentration ($n$), determined from room-temperature Hall effect measurements. It can be appreciated that $A$ varies within the $\approx (3\text{x}10^{-10} - 4\text{x}10^{-9})\,\Omega\,\text{cm}\,\text{K}^{-2}$ range when the carrier density ($n$) varies within the $\approx (3\text{x}10^{21} - 2\text{x}10^{22})\,\text{cm}^{-3}$ range (Supporting Information S3 and S4). In Figure 3a we also include the $A$ values reported in the literature for $Ca_{1-x}Sr_xVO_3$ polycrystalline samples $(4.2\text{x}10^{-10} - 9.1\text{x}10^{-10}\,\Omega\,\text{cm}\,\text{K}^{-2})$[7] as well as for SVO thin films $(2.5\text{x}10^{-10} - 5\text{x}10^{-10}\,\Omega\,\text{cm}\,\text{K}^{-2})$.[8–10] The close similarity of all available data is remarkable. To set in an appropriate context the large variation in magnitude of $A$, in Figure 3b we display the $A$ values for SVO and related oxide materials. We have selected oxides where the conduction band is mainly formed by partially occupied 3$d$-$t_{2g}$ orbitals, either intrinsic (SVO case) or obtained by electron doping of Mott insulators (e.g. Sr:GdTiO$_3$)[27] or band insulators (La:SrTiO$_3$, Nb:SrTiO$_3$ or SrTiO$_{3-x}$),[13,28,29] as first summarized by Mikheev et al.[30] and Stemmer et al.[11] It is remarkable that $A$ can be varied by about five orders of magnitude when the carrier concentration changes by about five orders of magnitude; we conclude that $A \approx n^{-1}$ over a very wide range of carrier concentration, as indicated by the dashed line in Figure 3a,b.

The temperature-dependent carrier mobility $\mu(T)$ ($\mu = (\rho\,n\,e)^{-1}$) was determined from $n(T)$, extracted from Hall effect measurements, and $\rho(T)$. Illustrative $n(T)$ and $\mu(T)$ data are shown in **Figure 4**a. It can be appreciated that $n(300\,\text{K}) \approx 2.5\text{x}10^{22}\,\text{cm}^{-3}$, reducing slightly ($\approx 12\,\%$) at 5 K. The mobility rapidly decays with increasing temperature. As shown in Figure 4b, $\mu^{-1}(T) \approx T^2$ which is similar to $\rho(T)$, implying that $\mu(T)$ governs $\rho(T)$. Therefore, we indistinctly refer to the quadratic temperature dependence of $\rho(T)$ or $\mu^{-1}(T)$.

Details of $\rho(T) = \rho_0 + A\,T^2$ fits can be appreciated in **Figure 5** where we show (dashed red) illustrative fits on samples having large $RRR = 9.6$ and small $RRR = 2.1$ (Figure 5a and 5b, respectively). It is obvious that $\rho(T)$ data of optimal films (Figure 5a) show a clear departure from the $T^2$ dependence, most noticeable at $T \lesssim 180$ K (dashed vertical line). Importantly, the deviation from $T^2$ becomes lest perceptible in films with lower $RRR$ (Figure 5b). Including a phonon-like $T^5$ term did not improve fits and did not affect the extracted $A$ parameter. Fits to other samples, residual fit differences and fits including additional $T^5$ terms are shown in Supporting Information S4.

To explore the origin of the departure of $\rho(T)$ from $T^2$ observed in the films with the largest $RRR$, we consider that the dynamics of the $d^1$ electron, moving in the narrow $3d$-$t_{2g}$ band within an ionic matrix ($O^{2-}/V^{4/5+}$), bears some similitude to polaronic motion. Indeed, it has been early suggested that polaronic carrier motion could be of relevance in itinerant-electron systems approaching the localized edge, including $CaVO_3$,[31] weakly-doped Ln:STO (where Ln is a lanthanide), orthorhombic manganites $La_{1-x}Ca_xMnO_3$,[32,33] $La_{0.7}Sr_{0.3}MnO_3$,[34] and $La_{2/3}(Ca_{1-x}Sr_x)_{1/3}MnO_3$,[5] or in doped $LaTiO_3$ and $NdTiO_3$.[35,36] Within this framework, the carrier mobility is determined by their coupling to some low energy phonons. The polaronic resistivity is given by:[33,37]

$$\rho(T) = \rho_0 + \frac{\hbar^2}{ne^2 a^2 t_\mathrm{p}} \cdot \frac{1}{\tau} \qquad (1)$$

where $t_\mathrm{p}$ is the hopping amplitude for polarons, $a$ the cell parameter, $n$ the carrier density and $\rho_0$ is the residual resistivity. $\tau^{-1}$ is the polaron relaxation rate, which is dictated by the $e$-$ph$ coupling to some phonon modes. In the simplest assumption of polaron coherent motion, in which a single optical phonon mode ($\hbar\omega_0$) dominates the polaron relaxation rate, $\tau^{-1}$ is given by:[33]

$$\frac{1}{\tau} = \frac{A_{\mathrm{e-ph}} \cdot \omega_0}{\sinh^2(\frac{\hbar\omega_0}{2k_\mathrm{B}T})} \qquad (2)$$

where $A_{\text{e-ph}}$ encapsulates the *e-ph* coupling strength and the effective mass of the polaron ($m^*_{\text{e-ph}}$) and $k_B$ is the Boltzmann constant. Assuming that $A_{\text{e-ph}}$ in Equation (2) is temperature independent, by combining all parameters, Equation (1) can be rewritten as:

$$\rho(T) = \rho_0 + \frac{A^*_{\text{e-ph}} \cdot \omega_0}{\sinh^2(\frac{\hbar \omega_0}{2 k_B T})} \tag{3}$$

where $A^*_{\text{e-ph}}$ is given by:

$$A^*_{\text{e-ph}} = \frac{\hbar^2}{n e^2 a^2 t_p} \cdot A_{\text{e-ph}} \tag{4}$$

In Figure 5a,b (dashed green lines), we include the fits of the $\rho(T)$ data using Equation (3), that allow to extract the phonon frequency $\omega_0$. We observe that data are well reproduced in all temperature range (5-300 K), including the low temperature region (< 180 K), where the quadratic fit failed. Residual fit differences are reduced by about 90% compared to $T^2$ fits and similar excellent fits have been obtained for all samples, and fitted parameters are robustly obtained irrespectively of fitting procedures (see Supporting Information S4).

Aiming at exploring changes of the phonon frequency ($\omega_0$) with lattice distortion, we first concentrate on fully strained films grown on substrates where epitaxial strain dictates different *c/a* ratios. One expects that phonons in SVO films are sensitive to tetragonal distortions of the VO$_6$ octahedra which can be quantified by the *c/a* ratio extracted from XRD. In Figure 5c,d we show the $\rho(T)$ data of SVO ($\approx$ 10 nm, *P(Ar)* = 0 mbar) films on STO, NGO and LAO. As observed, the polaronic model leads to excellent fits. The extracted $\hbar\omega_0$ values are around 20.6 meV (STO), 12.7 meV (NGO) and 10.7 meV (LAO), which indicates a clear softening under compressive distortion of the SVO lattice (from *c/a* < 1 to *c/a* > 1). **Figure 6**a (decorated sphere symbols) shows the observed $\hbar\omega_0$ and its variation with *c/a*. We also include in Figure 6a the phonon energy $\hbar\omega_0$ extracted from the fits of SVO films on optimally matched substrates (LSAT, NGO) grown at different pressures (0-0.3 mbar) and thicknesses (10-70 nm). For completeness, we have also included (light blue spheres) the $\hbar\omega_0$ data extracted from fits to digitized $\rho(T)$ data of films grown by hybrid-MBE (h-MBE),[9,25] which nicely fall on top of our data (fits are shown in Supporting Information Figure S4c).

In Figure 6b (left axis, full spheres), we explore a possible correlation between the *e-ph* coupling related $A^*_{e-ph}$ parameter and the phonon energy $\hbar\omega_0$. We also include data from literature[9,25] (light blue spheres) that follow the same $A^*_{e-ph}(\hbar\omega_0)$ trend: the *e-ph* coupling strength increases with $\hbar\omega_0$. To get an insight into the implications of data in Figure 6b, we evaluate $A_{e-ph}$ from $A^*_{e-ph}$ using Equation (4). A rough estimate of $A_{e-ph}$ can be obtained using $n \approx 2 \times 10^{22}$ cm$^{-3}$, $a \approx 4$ Å, and $t_p \approx 0.6$ eV [we use here the DFT calculated hopping integral for electrons (not polarons) in SVO[38]] and a typical value of $A^*_{e-ph}$ from Figure 6b ($\approx 1 \times 10^{-20}$ Ω m s). It turns out that $A_{e-ph} \approx 7$. In Figure 6b (right axis) we show the $A_{e-ph}$ values (empty circles) calculated using the actual *n*, the mean cell parameter $a = \sqrt[3]{V_{uc}}$ ($V_{uc}$ is the measured unit cell volume) and the experimental $A^*_{e-ph}$ values of all samples. These data provide a transparent view of the variations of the strength of the *e-ph* coupling with the phonon frequency. Within the polaronic framework, $A_{e-ph}$ is proportional to the effective mass of polarons, which data in Figure 6b, indicates that it becomes larger when phonons harden.

## 2.2. Temperature and magnetic field dependent magnetoresistance

In **Figure 7**a we display the magnetoresistance $MR(H) = [R(H)-R(H=0)]/R(H=0)$ of a representative SVO film (70 nm, NGO, $P(Ar) = 0.2$ mbar), recorded at various temperatures. The resistance $R(H,T)$ is measured with the magnetic field ***H*** perpendicular to the film plane and the current is transverse to ***H***. In a conventional metal $MR \approx (\omega_c \tau)^2 = (\mu \mu_0 H)^2$, where $\omega_c$ and $\tau$ are the cyclotron frequency and scattering time respectively, and $\mu$ and $\mu_0$ are the mobility and the vacuum permeability, respectively. Our experimental data show that *MR* of SVO is positive, parabolic on *H* and rather small (<2 % at 5 K), decreasing with increasing temperature. Fitting the parabolic $MR(H)$ gives $\mu \approx 30\,\text{cm}^2\,\text{V}^{-1}\,\text{s}^{-1}$ at 300 K, increasing up to $\mu \approx 160\,\text{cm}^2\,\text{V}^{-1}\,\text{s}^{-1}$ at 5 K (inset in Figure 7a). This increase of mobility upon cooling is consistent with Hall measurements (included also in inset in Figure 7a), although its value is somewhat smaller in the latter, as often observed.[39]

On the other hand, as stated by the Kohler's rule,[40] $MR(H)$ data recorded at different temperatures should be a unique function $F(x)$, with $x = [\mu_0 H/R(H=0)]$, as shown in Figure 7b. The Kohler's rule is expected to hold irrespectively on the carrier nature (e.g. correlated electrons or polarons) and scattering mechanism.[41]

Here it is relevant to emphasize that, in a non-magnetic system, $MR$ is non-zero only if different carriers participate in the transport, and this has to happen if the FS is constituted by interpenetrated cylinders as in the present case.

Indeed, first principles calculations (DFT) of the electronic structure of bulk SVO clearly indicate that the Fermi surface has a multiband character and is dominated by the $d_{xy}$, $d_{xz}$ and $d_{yz}$ orbitals, although a non-vanishing contribution $\approx 20\,\%$ (in terms of the density of states) of the oxygen $2p$ states is apparent. These orbitals determine a cylindrical FS along three-perpendicular directions (**Figure 8**). Calculations were performed for bulk SVO, using the cell parameters of tensely and compressively strained (10 nm) SVO films. They indicate that the fine details of the covalent mixing are somehow modified by strain, but the main features of the FS are fully preserved. Combination of three orbitals and the hybridization between the three FS sheets results in the obtained picture, where the outer sheet has a quasi-2D character and different carriers contribute to them, which can account for the observed magnetoresistance.

## 2.3. Seebeck coefficient

In **Figure 9,** we display the temperature dependence of the Seebeck coefficient $S(T)$ of an illustrative SVO film (70 nm, LSAT, $P(Ar) = 0.03$ mbar; similar data obtained for other samples are shown in Supporting Information S5). $S(T)$ is negative and increases linearly (in modulus) when increasing temperature, as expected for band transport of electrons in which $S(T)$ is given by:[15]

$$S(T) = \frac{-\pi^2 k_B^2}{3e} T \left[ \frac{g(E)}{n} + \frac{\partial}{\partial E} \ln[\tau(E)] \right]_{E=E_F} \quad (5)$$

where $g(E)$ is the density of states, $n$ the carrier density and $E_F$ the Fermi energy. $\tau(E)$ is an energy dependent scattering time, which in general can be written as $\tau \approx E^\alpha$ where $\alpha$ is related to the scattering mechanism. Within the simplest parabolic band approximation and for $\tau \approx E^{-1}$ ($\alpha = -1$), as deduced from optical measurements by Makino et al.,[42] $S(T)$ reduces to:

$$S(T) = \frac{-\pi^2 k_B^2}{6e} \frac{T}{E_F} \quad (6)$$

The slope $dS(T)/dT$ allows to extract $E_F$ and, using the carrier density $n$ deduced from Hall measurements ($n = 2.13 \times 10^{22}$ cm$^{-3}$), the transport effective mass $m^*$ can be determined. From Figure 9, $dS(T)/dT = -0.0211$ μV K$^{-2}$ and using Equation (6) we obtain $E_F = 0.58$ eV and $m^* \approx 4.8\, m_e$ ($m_e$ is the free electron mass). ARPES measurements indicate a Fermi energy at about $\approx 0.5$ eV[43] which is in excellent agreement with our Seebeck data.

Therefore, we conclude that transport in SVO is ruled by carriers having a large effective mass. Within the scope of the Brinkman-Rice model,[44] $m^* \approx 4.8\, m_e$ would indicate that the system is close to the metal-insulator transition. More elaborate ab-initio calculations for bulk SVO including only electronic correlation effects within dynamical mean field theory also find a significant mass enhancement.[45–47] This scenario has been proposed to account for the observation of a metal-insulator transition in ultrathin SVO films[8,10,48] and in irradiated SVO films.[49] However, as we argue in the following, the mass enhancement can also be contributed by *e-ph* coupling, as the polaronic fits suggest.

## 3. Discussion

We have first shown that the temperature-dependent $\rho(T)$ data of SVO films can be roughly described by either a quadratic $T^2$ dependence that we recall is commonly taken as a signature of *e-e* scattering in correlated electronic systems, or by a polaronic model. However, we have demonstrated that the quality of fits, performed in a wide temperature range (5-300 K), is significantly improved using a polaronic model. In the following, we carefully revise the implications of these findings.

**3.1. The quadratic temperature dependence of the resistivity**

The quadratic temperature dependence of $\rho(T)$ in metals is a common observation.[50] The *e-e* scattering rate is given by $1/\tau = B_{ee}(k_B T)^2/(\hbar E_F)$, where $B_{ee}$ is a dimensionless constant of order unity.[15] The electrical resistivity is given by $\rho = (m/ne^2)(1/\tau)$ and, if only the scattering time is temperature dependent, it follows that $\rho(T) \approx 1/\tau \approx T^2$. Accordingly, $\rho(T) \approx T^2$ can be viewed as a signature of FL.

For a spherical Fermi surface $E_F = (\hbar^2/2m)(3\pi^2 n)^{2/3}$ and $\rho(T) = (2m^2 k_B^2)/(e^2 \hbar^3 (3\pi^2)^{2/3} n^{5/3}) T^2 \approx n^{-5/3} T^2$.[51] Mobility data in Figure 4a,b as well as $\rho(T)$ in Figure 2d-f are roughly consistent with this $T^2$ dependence. However, the $A(n)$ data shown in Figure 3a,b is better described by $A \approx n^{-1}$ over a large range of carrier concentration, rather than $A \approx n^{-5/3}$ (dashed lines in Figure 3b). In the simplest view, $A \approx n^{-1}$ would imply that $1/\tau$ (and therefore $1/\mu$) would not depend on carrier density, which is at odds with expectations for a FL model. As recently emphasized by Stemmer et al.,[11,30] this discrepancy holds for a large number of oxides at the verge of a metal-insulator transition, although it is worth noticing that the predicted $A \approx n^{-5/3}$ dependence had been observed in other systems, such as TiS$_2$.[52]

The magnitude of the $A$ coefficient in $\rho(T) \approx A T^2$ signals another discrepancy. As shown in Figure 3a, we obtained $A \approx (1\times 10^{-10} - 1\times 10^{-9})\,\Omega\,\text{cm}\,\text{K}^{-2}$, which are similar values to those reported by Inoue et al. ($4.2\times 10^{-10}\,\Omega\,\text{cm}\,\text{K}^{-2}$).[7] Within the *e-e* scattering model, $A \equiv A_{ee}$ is given by:

$$A_{ee} = \frac{2m^2 k_B^2}{e^2 \hbar^3 (3\pi^2)^{2/3} n^{5/3}} \tag{7}$$

Using as typical parameters ($n = 2 \times 10^{22}$ cm$^{-3}$ and $m_{ee}^* \approx 4\, m_e$) one gets: $A \approx 1 \times 10^{-11}\,\Omega\,\text{cm}\,\text{K}^{-2}$, which is almost two orders of magnitude smaller than the measured values.

Summarizing, the observation that $A \approx n^{-1}$, suggesting that $\mu$ does not depend on $n$, and the severe discrepancy between measured and expected $A$ values, questions the *e-e* scattering in a FL scenario as the origin of the roughly quadratic temperature dependence of resistivity. Next, we investigate the role of the *e-ph* scattering.

### 3.2. Electron-phonon scattering in cylindrical Fermi surfaces

We first notice that the temperature dependence of $1/\tau$ is primarily due to scattering events with phonons. At high temperature, all phonon branches are equally populated and the phonon number increases linearly with $T$, therefore $1/\tau \approx T$ and $\rho(T) \approx T$, as frequently observed. However, at lower temperature the different phonon occupation of different phonon branches leads to a more complex situation, which becomes particularly remarkable for anisotropic Fermi surfaces. Indeed, it was long ago recognized that *e-ph* scattering in metals with cylindrical Fermi surfaces (e.g. Bi[14]) leads to $1/\tau \approx \rho(T) \approx T^2$ in some temperature range. As recently emphasized by Snyder et al.,[53,54] the topological 2D feature of the FS in some metallic oxides, can be at the origin of $1/\tau \approx \rho(T) \approx T^2$.[53,54] Indeed, a cylindrical Fermi surface of SVO has been observed by ARPES.[43,55–57] The recent observation of a similar $T^2$-dependent resistivity in Bi$_2$O$_2$Se oxyselenides, where the Fermi surface is an elongated ellipsoid,[12] may point to a common origin.

According to Kukkonen,[14] the electrical resistivity of metals with cylindrical Fermi surface display a genuine $T^2$ temperature dependence in a temperature region bounded by $T_p < T < T_k$. The temperature limits $T_p$ and $T_k$ are determined by the dimensions of the Fermi surface through the relation $T_p = 2\hbar v_S p_F / k_B$ and $T_k = 2\hbar v_S k_F / k_B$ where $p_F$ (resp. $k_F$) is the diameter (resp.

the height) of the cylindrical Fermi surface and $v_S$ the sound velocity.[14,58] In the SVO case, by approximating the FS to a cylinder, using $p_F \approx 0.5(\pi/a)$ and $k_F \approx (\pi/a)$, as deduced from ARPES experiments[43] and our calculations (Figure 8), and using the experimental transverse sound velocity $v_S \approx 4000$ m s$^{-1}$,[59] we get $T_p \approx 239$ K and $T_k \approx 478$ K. We notice that the estimated low-temperature limit ($\approx 239$ K) for $\rho \approx T^2$, is somewhat higher than the low temperature experimental bound for the quadratic $T^2$ term ($\approx 180$-$200$ K), below which a clear departure from $T^2$ is observed (Figure 5a,c). To what extent this discrepancy is related to limitations of the anisotropic scattering model of Kukkonen[14] (such as the assumption of a single band carriers and absence of phonon drag), or is linked to the present approximation of a (interpenetrated) cylindrical Fermi surface for SVO, remains to be solved.

Therefore, one could tentatively conclude that the $\rho(T) \approx T^2$ dependence may result from *e-ph* scattering in the quasi-2D cylindrical Fermi surface characteristic of 3$d$-$t_{2g}$ metal oxides. Seebeck and magnetoresistance data would be also compatible with this picture.

### 3.3. Polaronic transport

Landau first suggested the possibility for lattice distortions to trap electrons by means of an intrinsic modification of the lattice phonon-field induced by the electron itself. The resulting *e-ph* quasiparticle (the *polaron*) is a coupled *e-ph* system in which the polarization generated by the lattice distortions acts back on the electron, renormalizing its properties, for instance the effective mass.

At low temperature, (small) polarons may display a coherent band-like transport, where the phonon-mediated scattering rules their mobility.[60,61] As the phonon number decreases with temperature, the resistivity decreases upon lowering temperature and small polarons behave as heavy particles with effective mass $m^*_{e-ph}$. In this regime, $\rho(T)$ is given by Equations (1 − 4). As shown, data can be well reproduced by these expressions. As emphasized by Van der Marel et al.,[28] this $\rho(T)$ dependence is expected to hold for small polarons; but on the other hand, Devreese et al.[62] also signaled that, even in one of the most studied polaronic materials (Nb:SrTiO$_3$), the distinction between small and large polarons is not that sharp. We will not attempt to dig here into this distinction.

Instead, we note that in recent years, the small polaron scenario has been used to describe $\rho(T)$ of heavily doped manganites[33,63] or doped Mott insulators LaTiO$_3$[35,36] and NdTiO$_3$.[36] We focus now our attention on the relevant phonon energies ($\hbar\omega_0 \approx$ 5-25 meV $\approx$ 60-290 K) extracted from the fits of $\rho(T)$ of our films (Figure 6a) and the observed variation with the tetragonal distortion $c/a$. Preliminary calculations in SVO (Supporting Information S6) allow to identify phonons in this energy range that soften when increasing $c/a$, as observed experimentally (Figure 6a). Phonons within the same energy range were extracted from $\rho(T)$ data in manganites,[5,33,64] LaTiO$_3$[35] and Ba$_{1-x}$K$_x$BiO$_3$.[65,66] Therefore, we propose that similar phonons may govern the dynamics of *dressed* electrons in SVO films.

The *e-ph* coupling implies an enhanced effective mass ($m^*_{e-ph}$). Therefore, following Zhao et al.,[33] we identify in Equation (2), $A_{e-ph} \equiv \lambda$ where $\lambda = [(m^*_{e-ph}/m^*_{band}) - 1]$,[67,68] and we use $\lambda$ as a measure of *e-ph* coupling. It follows from data in Figure 6b, that $5 < \lambda < 10$, and accordingly, $6 < m^*_{e-ph}/m^*_{band} < 11$, depending on the tetragonality $c/a$ ratio Therefore, there is a dramatic enhancement of the electron effective mass via *e-ph* dressing. The large $m^*_{e-ph}/m^*_{band}$ effective mass derived from the polaronic fit is consistent with the large effective mass derived above from Seebeck data. Moreover, we notice that similar values have been reported in nickelates ($m^*/m^*_{band} \approx$ 6-7).[69]

Search for direct evidences of *e-ph* coupling in HTS was of paramount relevance in the quest for a microscopic mechanism for *e-e* pairing. Lanzara et al.[68] used angle-resolved photoemission spectroscopy (ARPES) to show that, in some cuprates, electrons experience an abrupt change of its velocity and scattering rate at some well-defined energy (50-80 meV) that was interpreted as a fingerprint of *e-ph* coupling. Interestingly, recent ARPES data in SVO,[43,55] showed similar features at ≈ 60 meV, that were also attributed to the coupling of electrons with these phonons. In principle, this conclusion would be in agreement with the polaronic model discussed here, although the significant difference on the energy of most relevant phonons for dc conductivity and ARPES remains to be elucidated.

Finally, we mention that Mirjolet et al.[23] and Zhang et al.[9] reported ellipsometric measurements to deduce the effective mass of carriers in SVO films, and obtained $m^*/m_e \approx$ 3-5 (depending on the substrates and growth conditions). Ellipsometric measurements have also been performed in some of the films of this manuscript, to determine the plasma frequency and consistent $m^*/m_e \approx 4.1$ values have been obtained (Supporting Information S8). It is also enlightening to notice that $m^*/m^*_{band}$ values extracted from specific heat coefficient ($\gamma$) and magnetic susceptibility ($\chi$) data of ceramic SVO samples differ by about a factor $R_W \approx 1.6$ (Wilson ratio), which also suggests that *e-ph* coupling to be relevant.[7]

To gain some perspective, it may be useful to point out that it has been recently shown that bosonic modes largely contribute to effective mass renormalization in SrRuO$_3$ oxide. SrRuO$_3$ is a metallic and ferromagnetic (< 150K) 4*d* system (Ru$^{4+}$: $d^4$: ($t_{2g}^{3\uparrow}$, $t_{2g}^{1\downarrow}$)) that had been commonly assumed to be a strongly correlated metallic system.[70,71] However, detailed calculations[72] have lately suggested correlations to be weaker than expected and ARPES data have provided strong evidence of *e-ph* coupling.[19] The fact that in both SrVO$_3$ and SrRuO$_3$, the itinerant electrons (3*d*-$t_{2g}^1$ and 4*d*-($t_{2g}^{3\uparrow}$, $t_{2g}^{1\downarrow}$), respectively) reside in a quasi-degenerate $t_{2g}$ band may be instrumental on the enhanced relevance of *e-ph* coupling.

We end by noticing that it has been recently reported that, beyond the original Kadowaki-Woods plot, there is a kind of universal link between the Fermi energy and the prefactor *A* of the $T^2$ resistivity, which persists across various Fermi liquids,[12] that remain to be explained. Our data also nicely fall within this scaling (see Supporting Information S8).

## 4. Conclusion

We have analysed transport properties of epitaxial SrVO$_3$ thin films, where V$^{4+}$ ions have a single 3$d^1$ electron in a $t_{2g}$ orbital triplet. First, we have shown that $\rho(T)$ displays roughly a $T^2$ dependence ($\rho(T) \approx \rho_0 + A\,T^2$) which is in agreement with earlier findings. However, the fit quality is unsatisfactory; the observed dependence of $A(n)$ is not that expected in a Fermi liquid and the magnitude of the $A$ coefficient differs by two orders of magnitude from expectations for $e$-$e$ scattering. This disconformity appears not only in SVO (intrinsic metal) but, as earlier pointed out, is shared by other conducting oxides (mainly doped semiconductors). We emphasize here that this discrepancy is common to oxides having a low occupation of narrow 3$d$-$t_{2g}$ bands. Two different scenarios have been considered to account for the available experimental data. We first note that the Fermi surface of these $3d^x$ (x $\leq$ 1) is mostly formed by three interpenetrated cylinders oriented along the three principal axis and thus the Fermi surface has a 2D character. As argued, the extreme anisotropy of the Fermi surface has a profound impact on the temperature dependence of the electron-phonon scattering and a $\rho(T) \approx T^2$ dependence was predicted in some temperature range, which is roughly in agreement with observations in SVO. For a $3d^1$ TMO, the FS includes up to three sheets, implying multiband conduction, which is consistent with the observed magnetoresistance.

Secondly, a polaronic (or vibronic) scenario has been explored. It has been shown that, assuming a single phonon mode ($\hbar\omega_0$) to be relevant for the $e$-$ph$ coupling, $\rho(T)$ can be excellently fitted in all temperature range (5-300 K). It is observed that the frequency of the relevant phonon can be tuned by the tetragonal distortion in SVO imposed by epitaxial strain and growth conditions. Moreover, it is found that $\hbar\omega_0$ and the $e$-$ph$ coupling strength ($\lambda$) both systematically increase under a tensile deformation of the lattice ($c/a$ < 1) and subsequently, the polaron effective mass, also increases.

In summary, the results indicate that *e-e* scattering in a FL alone does not account for the observed temperature dependence of resistivity, mobility and magnetoresistance in these transition metal oxides. Instead, other ingredients need to be invoked. The cylindrical 2D-like nature of the Fermi surface of SVO and the *e-ph* coupling giving rise to a polaronic transport, appear to be necessary ingredients to account for available transport, calorimetric and spectroscopic data. SVO and presumably other $3d^x$ (x ≤ 1) oxides may share a similar *e-ph* coupling that has remained largely unexplored. These findings may have some practical consequences. To mention one, in the search for transparent conducting oxides, where focus was on correlated systems, the present findings suggest that enhanced *e-ph* coupling could be an efficient tool to bring the plasma frequency to the infrared region.

## 5. Experimental Section/Methods

*Samples Preparation*: SVO films of thickness 10-70 nm have been grown on single crystalline SrTiO$_3$ (STO), (LaAlO$_3$)$_{0.3}$(Sr$_2$TaAlO$_6$)$_{0.7}$ (LSAT), NdGaO$_3$ (NGO) and LaAlO$_3$ (LAO) substrates by pulsed laser deposition (PLD), as described in detail elsewhere.[23] Bulk SVO is cubic with cell parameter $a_{SVO}$ = 3.842 Å. The (pseudo)cubic cell parameters of the substrates ($a_S$) are: 3.905 Å (STO), 3.868 Å (LSAT), 3.863 Å (NGO) and 3.791 Å (LAO). The corresponding structural mismatch parameters (defined as $f = (a_S - a_{SVO})/a_S$) are: +1.59 %, +0.65 %, +0.52 % and −1.37 %, respectively. It is known that the residual resistivity ($\rho_0 \approx \rho(5$ K)) and the residual resistivity ratio ($RRR = \rho(300$ K)$/\rho(5$ K)) both depend on growth conditions, namely gas pressure. Consequently, films have been grown at various Ar pressure ($P(Ar) = 0 - 0.3$ mbar).[24] Earlier experiments have shown that optimal SVO films can be grown at 750 °C, and this temperature has been kept fixed for all films reported here.[23,24] To further modulate the carrier density, another series of films has been deposited at various oxygen pressure $PO_2$ (from the base pressure of the chamber, i.e. ≈ 4x10$^{-7}$ mbar, to 2x10$^{-5}$ mbar). For growth details, see Mirjolet et al.[23]

*Structural Characterization*: X-ray diffraction (XRD) techniques (*θ-2θ* symmetric scans and reciprocal space maps) have been used to determine the in-plane (*a*) and out-of-plane (*c*) cell parameters of the epitaxial films. Illustrative XRD patterns, extracted (*a,c*) parameters and tetragonal distortion (*c/a* ratio) are shown in Supporting Information Figure S1. The film thickness was determined by X-ray reflectivity (XRR) and/or by fitting of the Laue fringes (when present). In agreement with previous reports,[23,24] it was found that the *c/a* ratio in SVO thin films differs from bulk SVO due to epitaxial strain and growth-induced defects. Indeed, in SVO films grown on LSAT or NGO, the tetragonal *c/a* ratio varies systematically (0.986 < *c/a* < 1.006) with *P(Ar)*, typically increasing when reducing *P(Ar)* as a result of the out-of-plane unit cell expansion in films grown at the lowest pressures.

*Electrical Characterization*: Electrical measurements on the films have been performed by using four-probe Van der Pauw contact configuration. Hall effect and magnetoresistance have been recorded by using magnetic fields up to $\pm 9$ T applied perpendicular to sample surface. Illustrative Hall effect measurements are shown in Supporting Information Figure S3. In agreement with earlier findings, the carrier density extracted from Hall effect, assuming a single band, varies within the $n \approx (1.8 \times 10^{22} - 2.6 \times 10^{22})$ cm$^{-3}$ range depending on film thickness, substrate and growth conditions. The carrier density roughly coincides with the expected carrier density of SVO (1 electron/unit cell). The carrier density of all films is included in Supporting Information S3. For the measurements of the Seebeck coefficient, two Cr/Pt (5/50 nm) lines with four contacts (1 mm x 50 μm, 2 mm apart) were deposited by optical lithography on top of the film. After deposition, a current of 1 mA was driven through the Pt lines at room temperature during 5 min to favor the recrystallization of Pt. The temperature dependence of the resistivity of each Pt line was measured in several cooling/heating ramps, until obtaining a reproducible result. The Pt resistivity was finally recorded on a heating ramp at 0.5 K min$^{-1}$ and the recorded values were used as local thermometers. Before each Seebeck measurement, the sample was stabilized at the base temperature for at least 15 min to ensure the absence of spurious thermal gradients that could influence the determination of the intrinsic Seebeck voltage. Different currents were injected through the heater until a constant temperature difference between the Pt thermometers was achieved. The voltage between the Pt lines is measured at the same position with a switch. Fitting the voltage vs temperature difference provides an accurate measurement of the Seebeck coefficient. An example of measurement procedure is given in the Supporting Information Figure S5.

*Optical Measurements:* Spectroscopic ellipsometry measurements (Figure S7 in the Supporting Information S7) were done in the far-infrared (FIR) and mid-infrared (MIR) using an IR spectroscopic ellipsometer, based on Bruker Vertex 70v FTIR spectrometer, similar to one described in Ref.[73] Near-infrared to UV part of the spectrum was determined with Woollam VASE (variable angle of incidence spectroscopic ellipsometer).

*First Principles Calculations:* The electronic structure of bulk $SrVO_3$ was calculated using density functional theory (DFT),[74,75] as available in the all-electron full-potential localized orbitals (FPLO) basis set code.[76] The generalized-gradient approximation[77] was chosen for the exchange-correlation energy and the summation in the Brillouin zone was performed on the (20x20x20) *k*-mesh. The Fermi surface was determined based on the band energies calculated on the (50x50x50) *k*-mesh. The structural parameters of bulk $SrVO_3$ were taken from reported data for the bulk material (lattice parameter of 3.842 Å) and the measured (*a, c*) parameters of the films were used when appropriate. In order to calculate the phonon properties, we used the density functional perturbation theory as available in the Quantum Espresso package.[78] In the first step, the electronic structure was determined self-consistently using the density functional theory[74] within the PBE parametrization of the generalized-gradient approximation. The electronic wavefunctions were represented by plane waves with an energy cutoff of 80 Ry. For the electronic density, larger cutoff of 320 Ry was used. The smearing for the electronic occupations was set to 0.02 Ry and the integration in the Brillouin zone was done on the Gamma-centered (10x10x10) *k*-mesh. The calculation of the electron-phonon interaction was based on a denser (20x20x20) $\boldsymbol{k}$-mesh and the phonon modes at the Gamma point were determined with the threshold $10^{-16}$.

**Acknowledgements**
Financial support from the Spanish Ministry of Science, Innovation and Universities, through the "Severo Ochoa" Programme for Centres of Excellence in R&D FUNFUTURE (CEX2019-000917-S) and MAT2017-85232-R (AEI/FEDER, EU) projects, from Generalitat de Catalunya (2017 SGR 1377), and from Deutsche Forschungsgemeinschaft (DFG), through TRR 288-422213477 (A05), is acknowledged. The authors would also like to acknowledge the Computer Center of the Goethe University Frankfurt for providing the computational resources. The work of M.M. has been done as a part of the Ph.D. program in Physics at Universitat Autònoma de Barcelona, and was financially supported by the Spanish Ministry of Science, Innovation and Universities (BES-2015-075223). The contribution of Clemens Lindermeir to the first stages of data analysis is acknowledged.

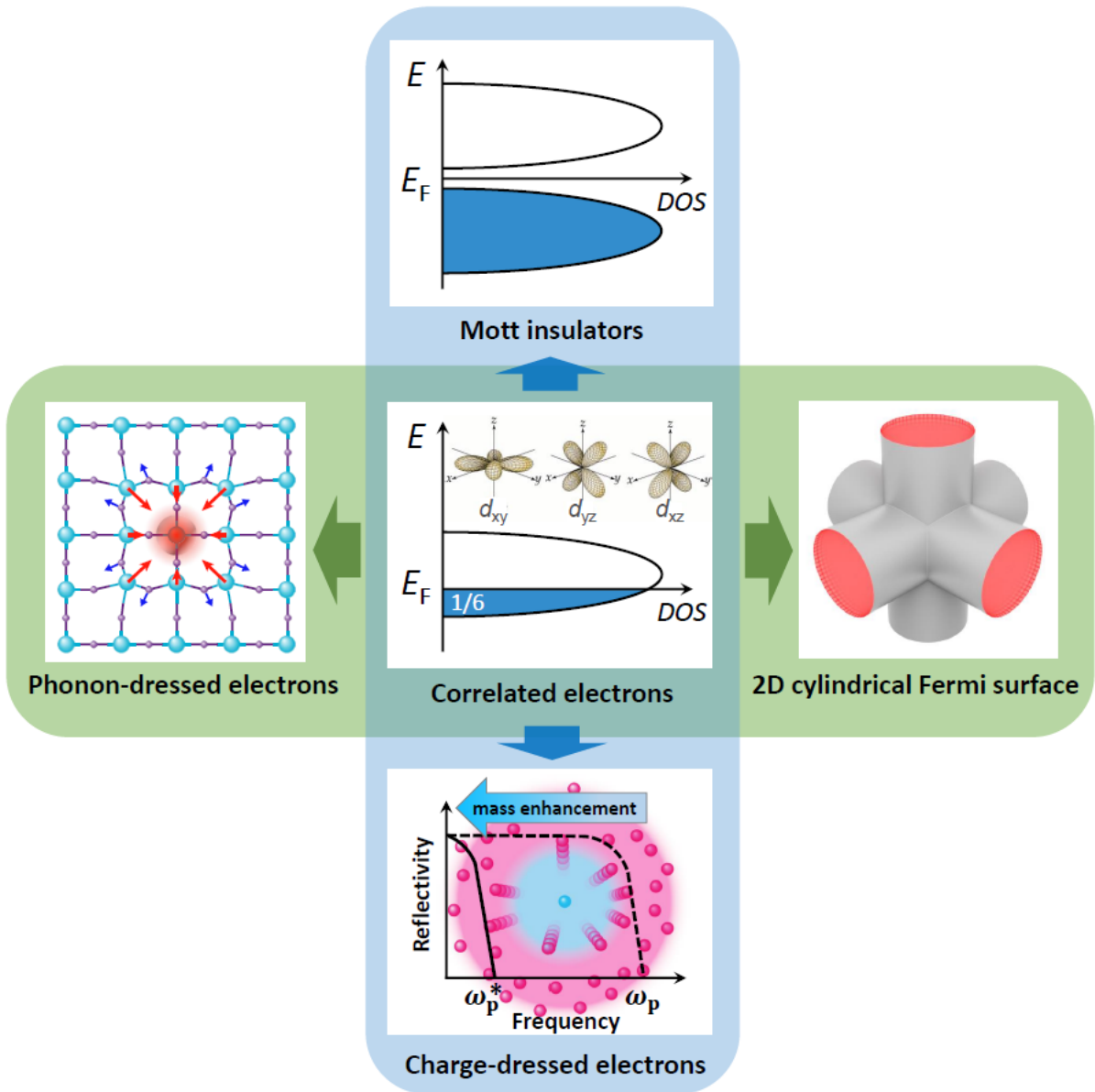

**Figure 1.** (Center) Partially occupied $t_{2g}$ ($d_{xy}$, $d_{xz}$, $d_{yz}$) orbitals, (1/6) in the sketch, are responsible for metallic behaviour. Electron-electron correlations may increase carrier effective mass ($m^*_{ee}$) and, among other consequences, reduce the plasma frequency (bottom illustration) or, eventually, open a Mott gap (top illustration). In cubic metal-oxide surrounding, the symmetry of the ($d_{xy}$, $d_{xz}$, $d_{yz}$) orbitals produces a quasi 2D cylindrical Fermi surface (right). Charge carriers in an ionic lattice are dressed with a lattice polarization cloud (left illustration), enhancing also the effective mass ($m^*_{e-ph}$).

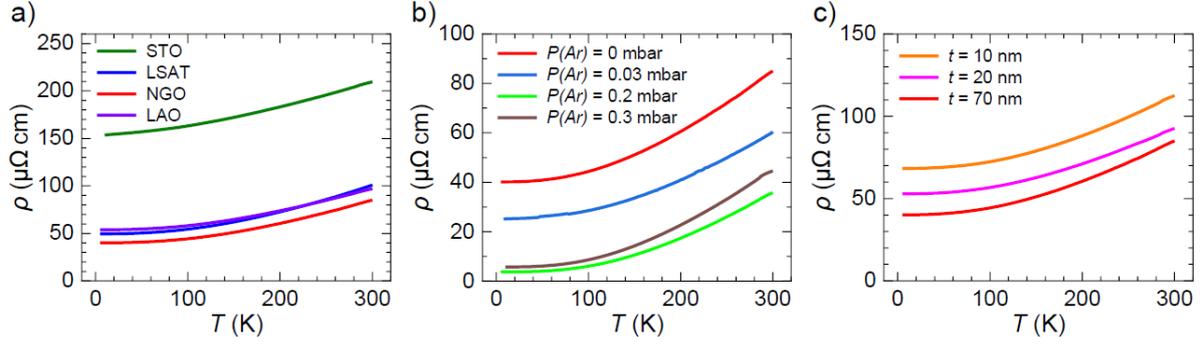

**Figure 2.** Temperature dependence of $\rho(T)$ of selected series of SVO samples, plotted vs $T$. a) Films of 70 nm gown at $P(Ar)=0$ mbar on various substrates; b) Films of 70 nm on NGO grown at various $P(Ar)$. c) Films of different thickness, on NGO grown at $P(Ar)=0$ mbar.

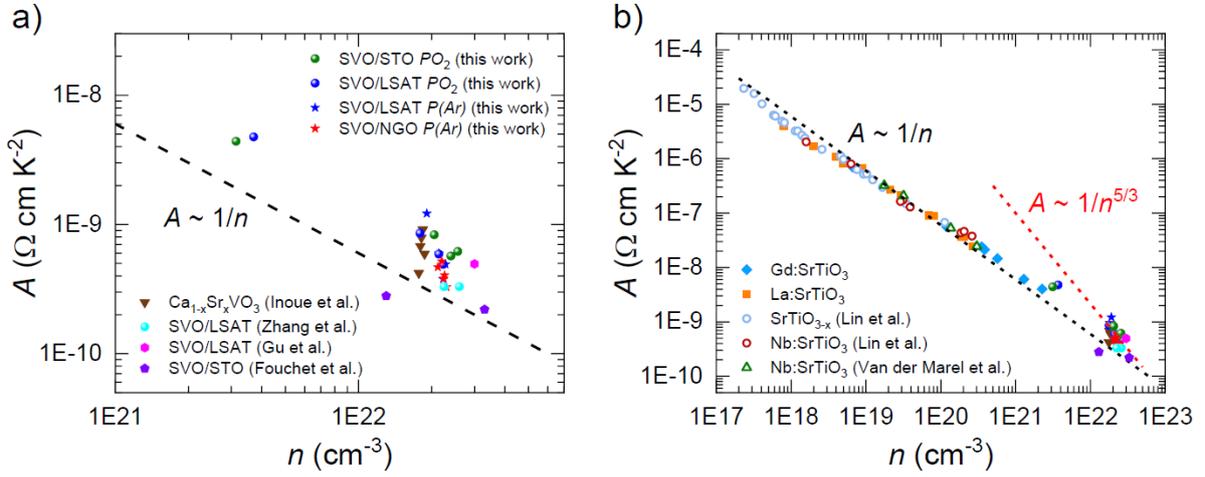

**Figure 3.** a) Coefficient $A$ of $\rho(T)=\rho_0+A\,T^2$ vs carrier density ($n$), as extracted from Hall measurements, for some SVO films. Data from literature references[7–10] are also included. b) Comparison of present $A$ values with available literature values for SVO films and related metallic oxides (figure adapted from Mikheev et al.[30] and Stemmer et al.[11], including data of Gd:SrTiO$_3$ from Moetakef et al.[27], La:SrTiO$_3$ from Cain et al.[29], Nb:SrTiO$_3$ from Van der Marel et al.[28], and SrTiO$_{3-x}$ and Nb:SrTiO$_3$ from Lin et al.[13]). Dashed lines indicate $A \approx n^\alpha$ power dependences with $\alpha = -1$ and $-5/3$.

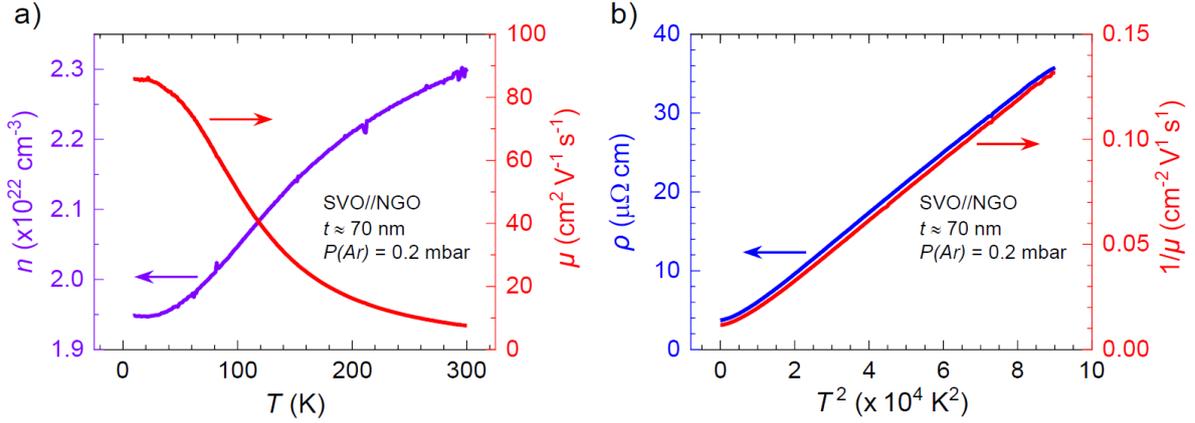

**Figure 4.** a) Temperature-dependent carrier concentration (left axis) and mobility (right axis) of an illustrative SVO film (70 nm thick, deposited on NGO at $P(Ar) = 0.2$ mbar). b) Resistivity (left axis) and inverse mobility (right axis) vs $T^2$ of the same film.

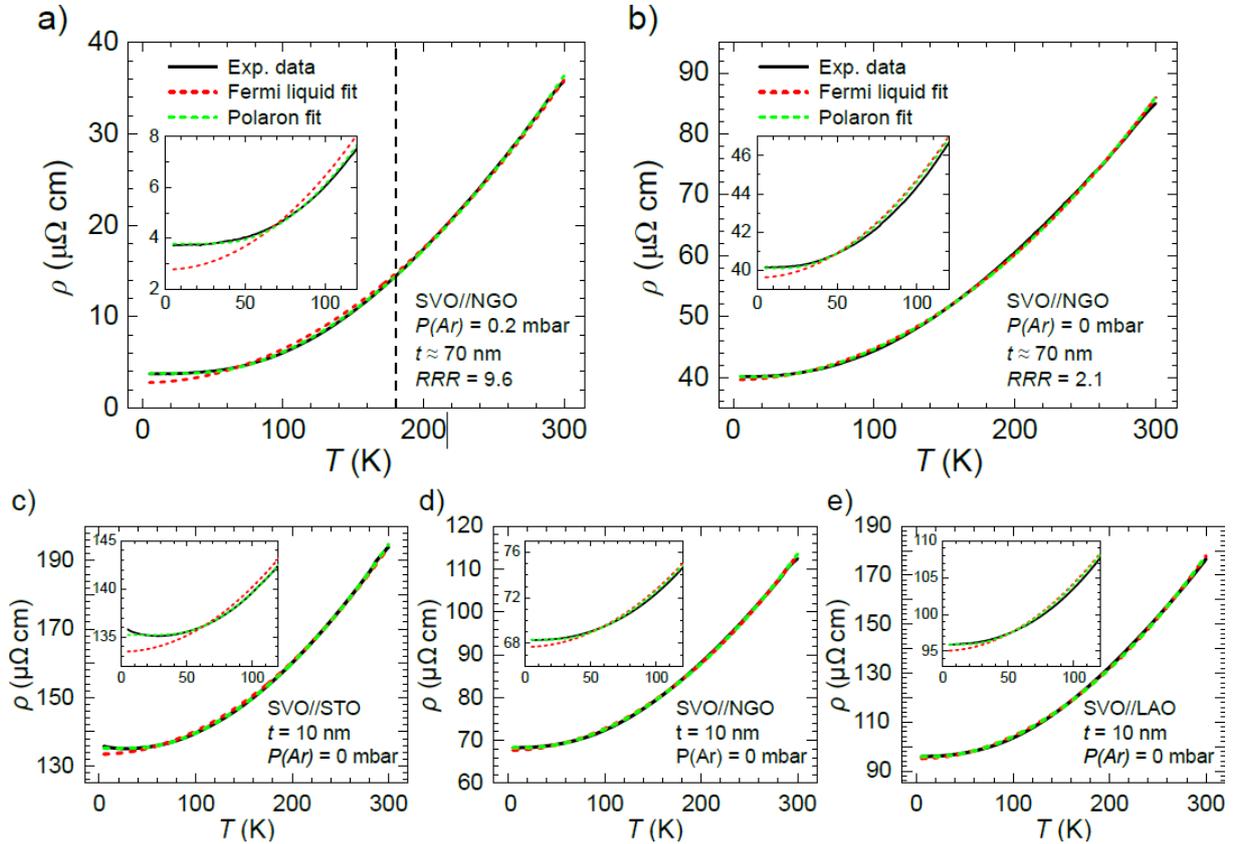

**Figure 5.** a,b) Illustrative $\rho(T)$ data of SVO films on NGO substrates (70 nm, grown at different $P(Ar)$) largely differing in *RRR*, as indicated. c,d,e) $\rho(T)$ data of SVO films (10 nm, $P(Ar) = 0$ mbar) grown on substrates imposing tensile (STO, NGO) or compressive strain (LAO) on SVO. The dashed red line is the fit to the $T^2$ dependence. Dashed vertical black lines indicate the temperature where noticeable departure from the quadratic $T$-dependence is observed. The dashed green line is the fit to the polaronic model. Insets are zooms of the low temperature region.

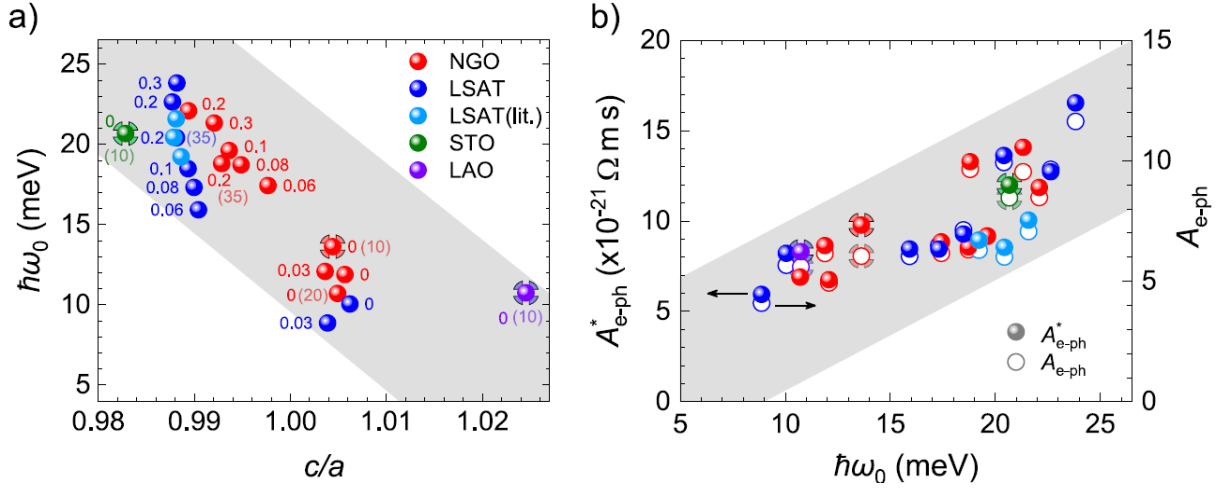

**Figure 6.** a) Dependence of the phonon energy ($\hbar\omega_0$) extracted from $\rho(T)$ using the polaronic model (Equation (3)) of SVO films as a function of their tetragonal cell distortion *(c/a)*. Films were grown on NGO (red spheres), LSAT (blue), STO (green) and LAO (violet), at different *P(Ar)* (as indicated by the labels; units are mbar). All films are 70 nm, except some few as indicated by the additional label in parenthesis. b) Left axis (full spheres): Dependence of $A^*_{e-ph}$ on the phonon energy ($\hbar\omega_0$). Right axis (empty circles): Electron-phonon coupling parameter $A_{e-ph}$ calculated from the experimental $A^*_{e-ph}$ as described in the text (Equation (4)). In (a,b) we also include (light blue spheres) the corresponding data points extracted from reported resistivity data of SVO films deposited on LSAT by h-MBE, by Zhang et al. (20 and 45 nm thick)[9] and by Moyer et al. (50 nm thick).[25] Errors bars for fit-extracted $\omega_0$ and $A^*_{e-ph}$ in (a,b) are smaller than the symbol size (Supporting Information S4).

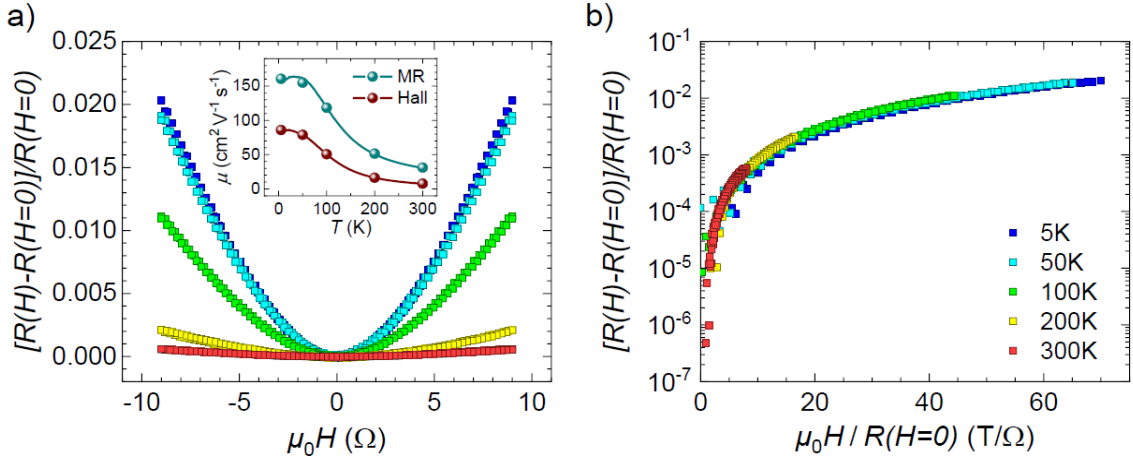

**Figure 7.** a) Magnetoresistance $MR = [R(H) - R(H=0)]/R(H)$ of SVO film (70 nm thick; deposited on NGO substrate; $P(Ar) = 0.2$ mbar) recorded with magnetic field perpendicular to the film surface, at various temperatures (5, 50, 100, 200 and 300 K). Data display a clear parabolic dependence $MR \approx H^2$. Inset: Carrier mobility extracted from Hall effect and magnetoresistance measurements as indicated. b) Kohler plot of $MR(H)$ measured at 5, 50, 100, 200 and 300 K, illustrating the expected collapsing.

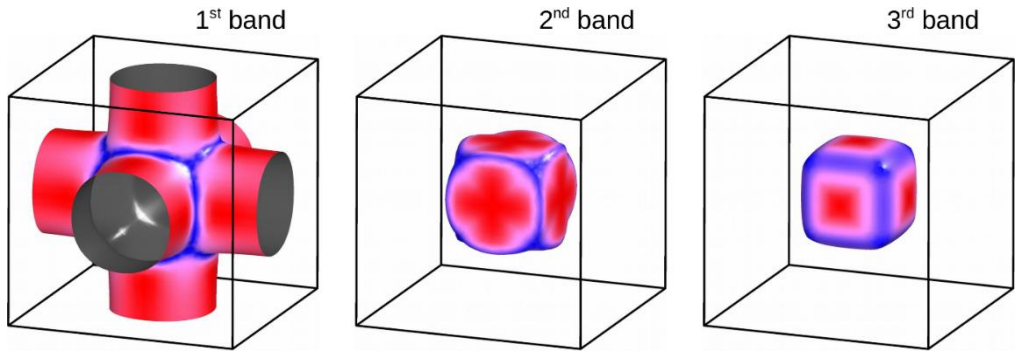

**Figure 8.** Fermi surface of cubic SVO determined from first principles. Three sheets of the Fermi surface are shown. The aspect ratio of the quasi-cylindrical 1$^{st}$ band is 1:2.1.

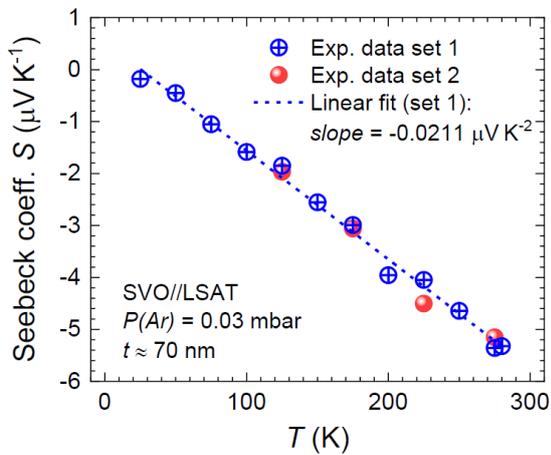

**Figure 9.** Temperature dependence of the Seebeck coefficient $S$ measured on a SVO film (70 nm thick; deposited on LSAT). Open blue symbols (data set 1) are data measured when ramping the temperature. Solid red symbols (data set 2) are data determined in a different temperature run, by setting the temperature before measuring the Seebeck voltage.

# Supporting Information

**Electron-Phonon Coupling and Electron-Phonon Scattering in SrVO₃**

*Mathieu Mirjolet, Francisco Rivadulla, Premysl Marsik, Vladislav Borisov, Roser Valentí, and Josep Fontcuberta**

**Supporting Information S1: Structural and topographic morphological data.**

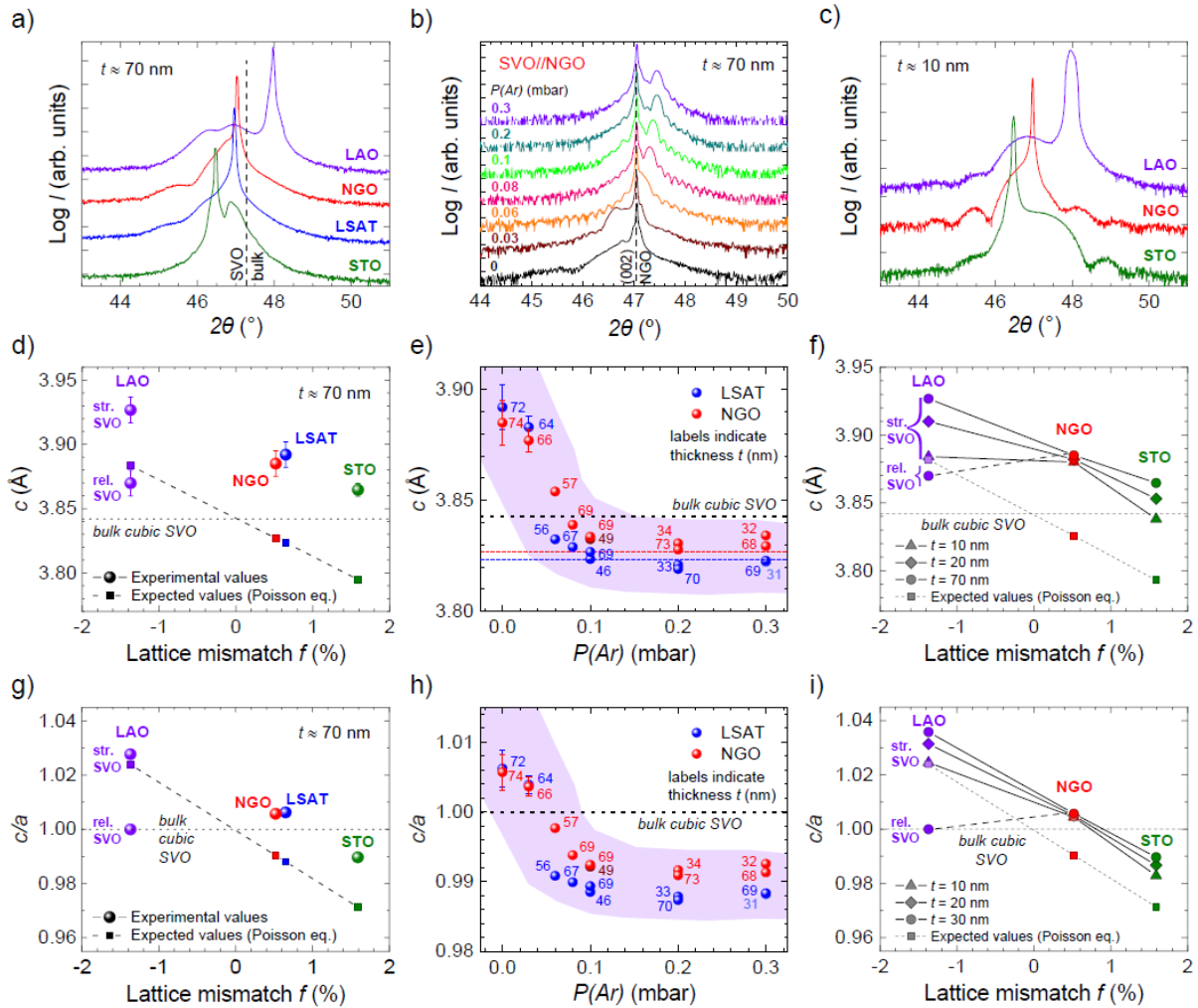

**Figure S1a:** (a-c) Illustrative X-ray diffraction patterns for most of SVO films of this study. From left to right column: strain series (different substrates, at constant thickness of about 70 nm), $P(Ar)$ series on LSAT and NGO, and thickness series (10, 20 and 70 nm) on various substrates. (d-f) Extracted $c$-axis parameters for the corresponding samples. To minimize the measurement error, the $c$-axis was extrapolated by Nelson-Riley method using the (00$l$) family ($l = \{1; 2; 3; 4\}$) of diffraction peaks. From our reciprocal space maps[1] all films deposited on NGO, LSAT and STO (tensile strain) were fully strained ($a_{SVO} = a_S$). Thick SVO films (70 nm) deposited on LAO (compressive strain) show strain relaxation, while thinner ones (10-20 nm) are nearly fully strained. (g-i) Corresponding tetragonal distortion ($c/a$ ratio).

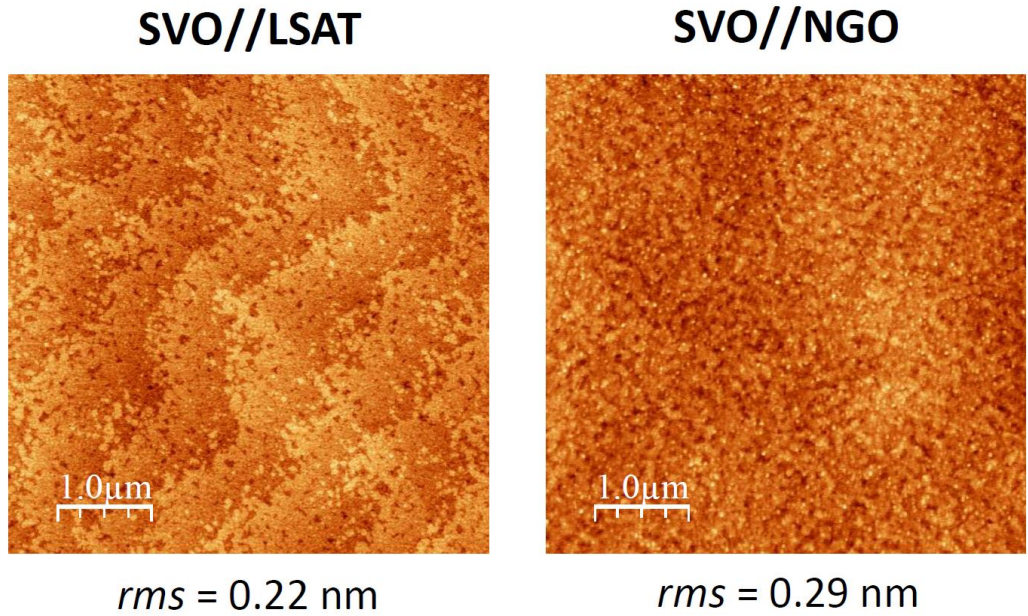

**Figure S1b:** Illustrative AFM topographic images of SVO films grown on LSAT (left) and NGO (right), at $P(Ar) = 0.03$ mbar. Films were about 70 nm thick. Image size is 5 μm x 5 μm.

**Supporting Information S2: Carrier density, carrier mobility, and residual resistivity ratio.**

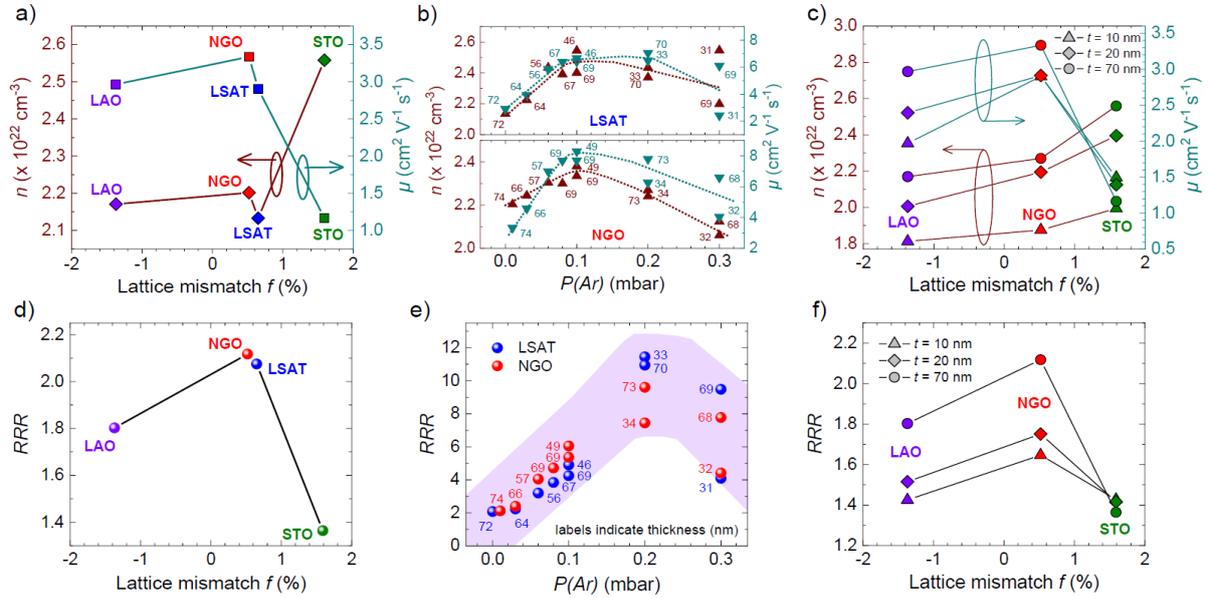

**Figure S2:** Transport data for most SVO films of this study. From left to right column: strain series (different substrates), $P(Ar)$ series on LSAT and NGO, and thickness series (10, 20 and 70 nm) on various substrates. Upper panels show the room-temperature carrier density $n$ and mobility $\mu$. Lower panels show residual resistivity ratio ($RRR$) of the corresponding samples.

**Supporting Information S3: Illustrative Hall effect measurements.**

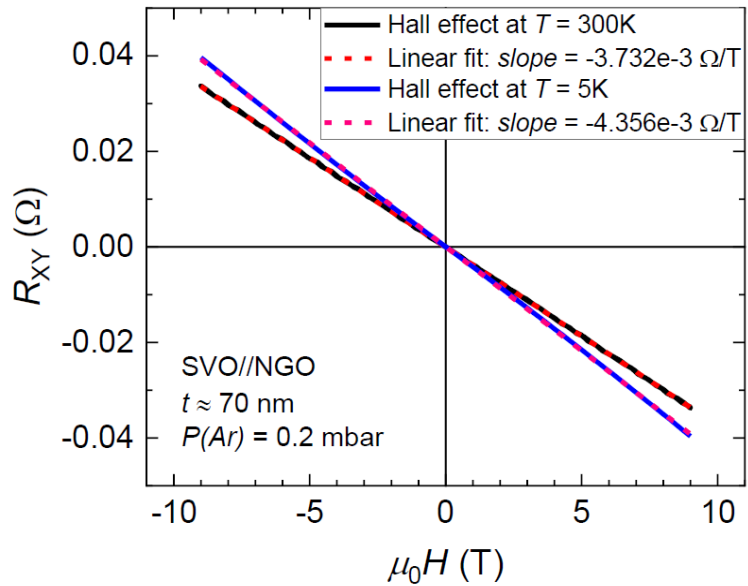

**Figure S3:** Illustrative Hall effect measurements at 300 K and 5 K of a 70 nm thick SVO film deposited on NGO substrate, at $P(Ar) = 0.2$ mbar.

**Supporting Information S4: Fits of resistivity data to a quadratic temperature dependence and polaronic models.**

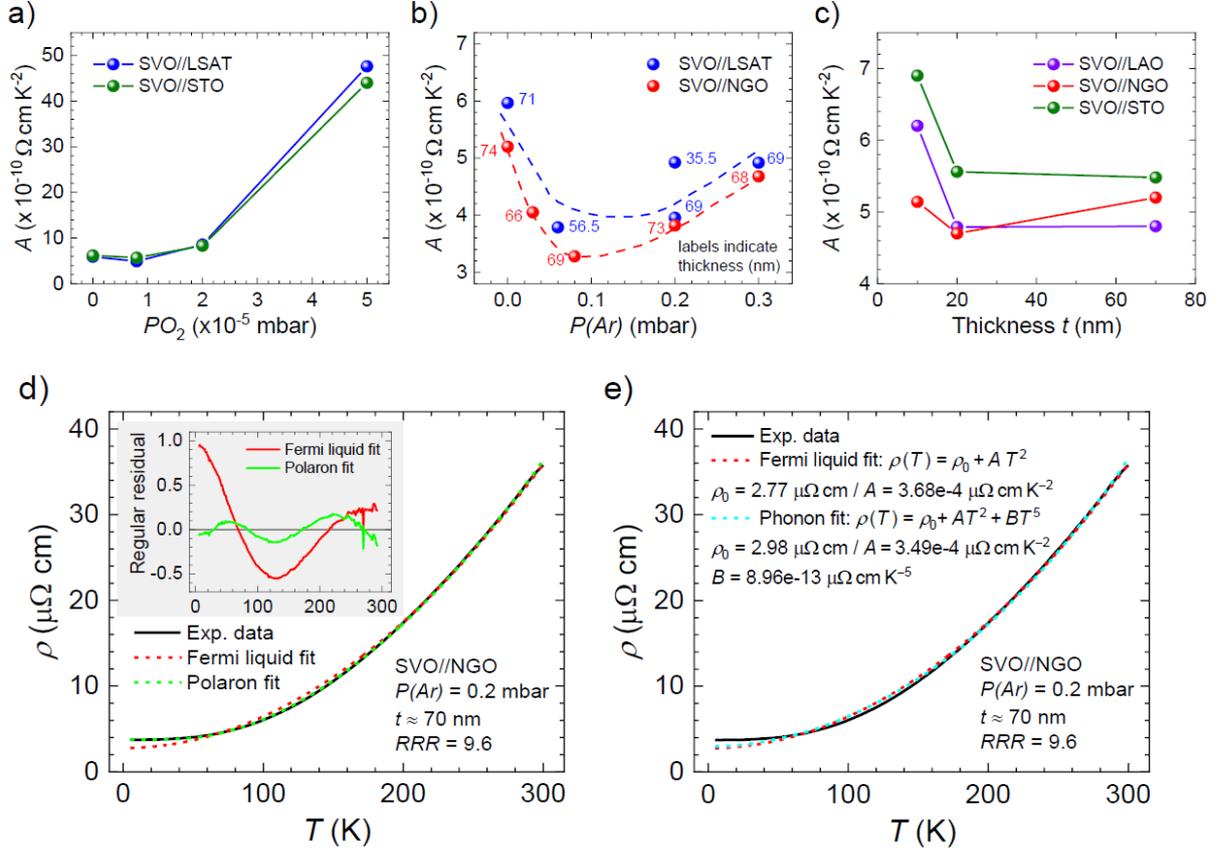

**Figure S4a:** Compilation of $A$ values extracted from the fitting function $\rho(T)=\rho_0+AT^2$. a) $PO_2$ series on STO and LSAT substrates. b) $P(Ar)$ series on LSAT and NGO substrates. c) Thickness series on LAO, NGO and STO substrates. d) Illustrative results of the fit of $\rho(T)$ using the $T^2$-dependence and the polaronic model. Inset show the residual difference $[\rho_{exp}-\rho_{fit}]$. Notice that the residual difference at $T<180\text{-}200$ K becomes about a factor 10 larger in the quadratic model. e) Comparison of the $T^2$ and $AT^2+BT^5$ fits together with the polaronic model, for an illustrative film.

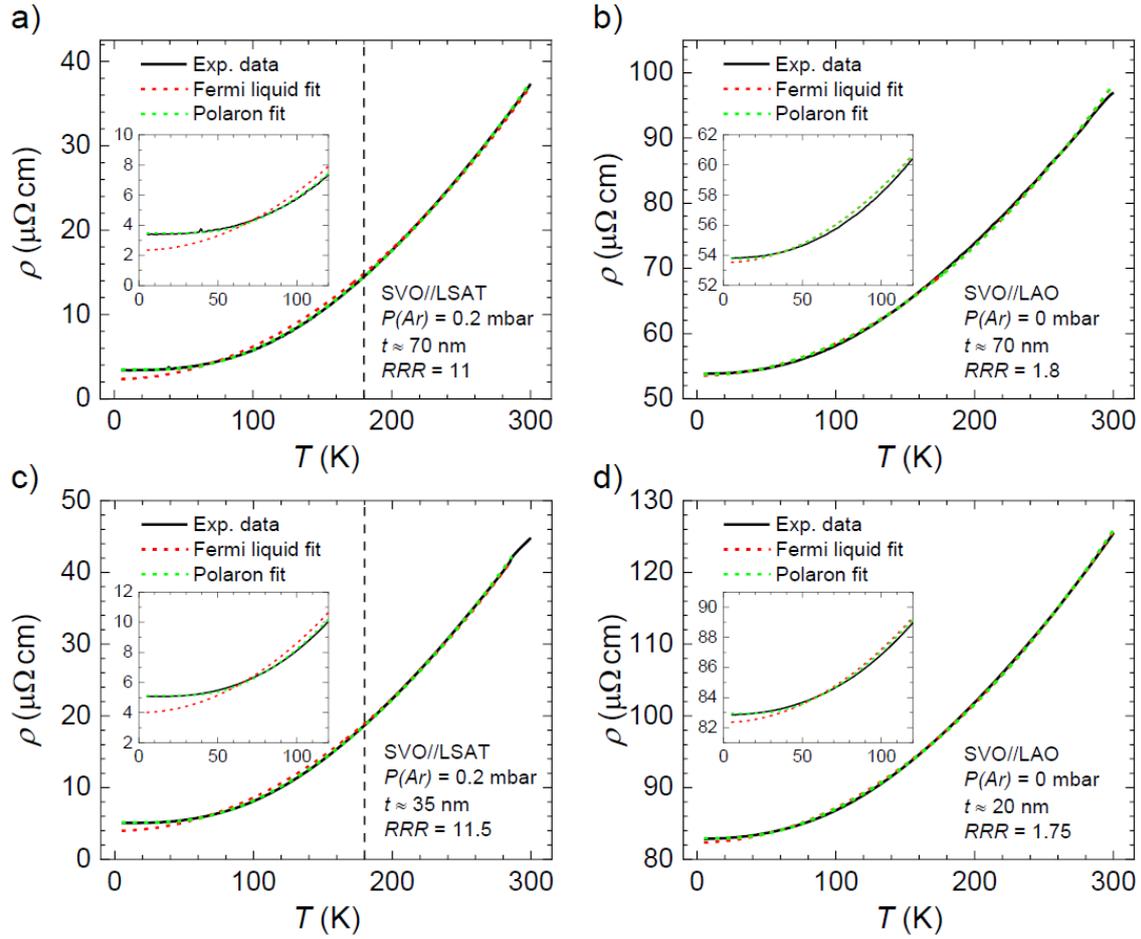

**Figure S4b:** Additional $\rho(T)$ data of SVO films together with Fermi liquid fit (red dashed curve) and polaron model fit (green dashed curve). Left panels show SVO of thickness: a) 70 nm and c) 35 nm thick film; deposited on LSAT at $P(Ar) = 0.2$ mbar. Right panels show SVO film of thickness: b) 70 nm and d) 20 nm; deposited on LAO at $P(Ar) = 0$ mbar. Insets are zooms of the low temperature region where the Fermi liquid fits show highest discrepancy.

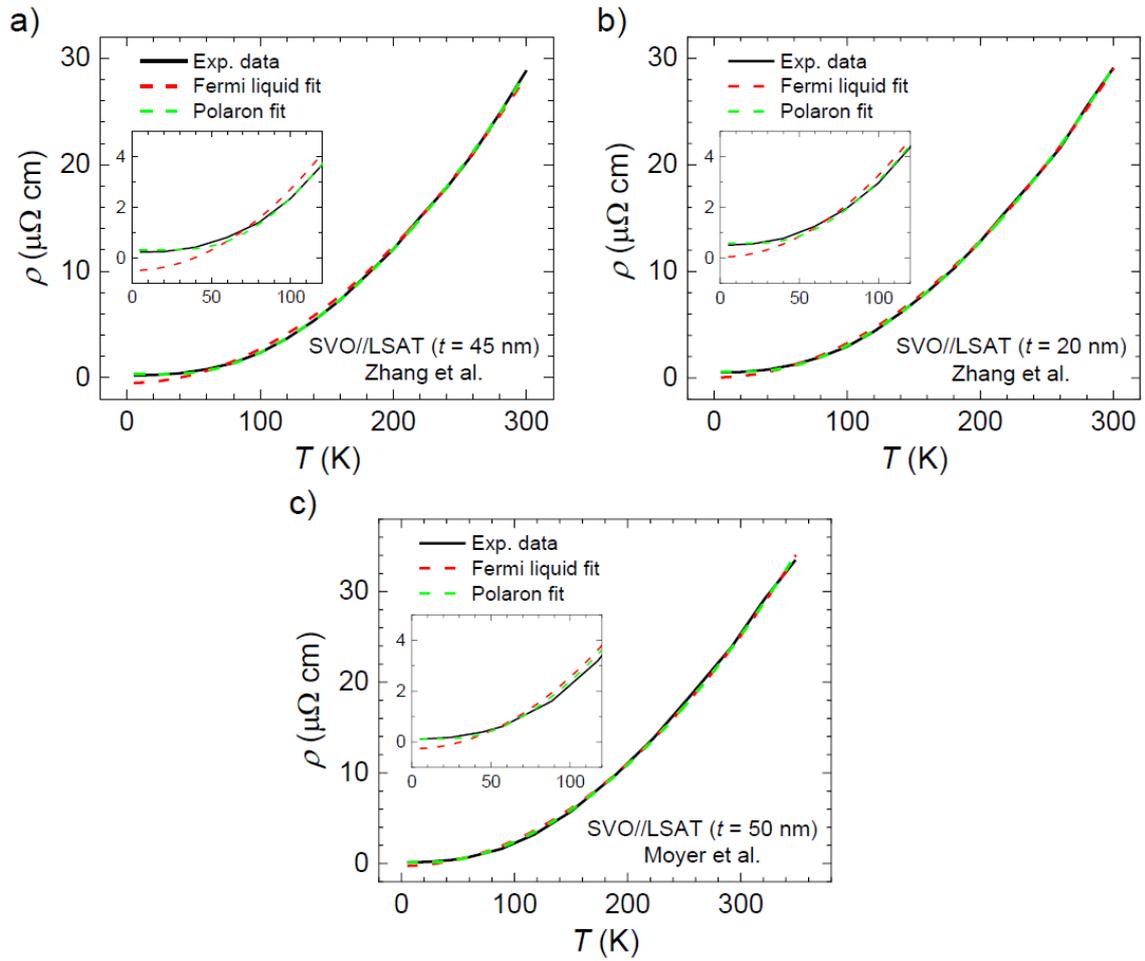

**Figure S4c:** Literature ρ(T) data of SVO films grown on LSAT by hybrid-MBE, together with Fermi liquid fit (red dashed curve) and polaron model fit (green dashed curve). Data in (a,b) represent digitized data of 45 and 20 nm thick films, respectively, from Zhang et al.[2] Data in (c) are taken from Moyer et al.[3] (50 nm thick). Insets are zooms of the low temperature region where the Fermi liquid fits show highest discrepancy.

**Fitting parameters:**

**Table S4-I:** Fitting parameters for Fermi liquid (constrained and unconstrained) and polaronic fits, for some illustrative SVO films: SVO films (10 nm) on STO, NGO and LAO (data are shown in Figure 5). Notice the errors of parameters are < 6%.

| Sample | Fit | $R^2$ | $A_{ee}$ [Ω cm K$^{-2}$] | $A^*_{e-ph}$ [Ω m s] | $\hbar\omega_0$ [meV] |
|---|---|---|---|---|---|
| SVO//STO (Fig. 5c) | Fermi liquid (fixed $\rho_0$) | 0.99**314** | 6.35e-10 ± 2.14e-12 | x | x |
| | Fermi liquid | 0.99**860** | 6.72e-10 ± 1.46e-12 | x | x |
| | Polaronic | 0.99**995** | x | 1.20e-20 ± 1.29e-23 | 20.6 ± 0.11 |
| SVO//NGO (Fig. 5d) | Fermi liquid (fixed $\rho_0$) | 0.99**867** | 4.955e-10 ± 7.12e-13 | x | x |
| | Fermi liquid | 0.99**962** | 5.07e-10 ± 5.74e-13 | x | x |
| | Polaronic | 0.99**980** | x | 9.78e-21 ± 2.77e-22 | 12.74 ± 0.35 |
| SVO//LAO (Fig. 5e) | Fermi liquid (fixed $\rho_0$) | 0.99**899** | 9.04e-10 ± 1.12e-12 | x | x |
| | Fermi liquid | 0.99**954** | 9.20e-10 ± 1.15e-12 | x | x |
| | Polaronic | 0.99**962** | x | 8.30e-21 ± 4.58e-22 | 10.75 ± 0.58 |

**Table S4-II:** Comparison of fitted parameters depending on input parameters for the polaronic fit. It can be appreciated that the input parameters can varied by about 3 orders of magnitude but the fitted values are virtually identical.

| Sample | Initial $\hbar\omega_0$ [meV] | Initial $A^*_{e-ph}$ [Ω m s] | Reduced chi-squared | $R^2$ | $\hbar\omega_0$ [meV] | Dependency | $A^*_{e-ph}$ [Ω m s] | Dependency |
|---|---|---|---|---|---|---|---|---|
| SVO//NGO (Fig. 5d) | **10** | **1e-20** | 0.0374 | 0.9998 | 12.74 | 0.9996 | 9.78e-21 | 0.9996 |
| | 100 | 1e-20 | 0.0374 | 0.9998 | 12.74 | 0.9996 | 9.78e-21 | 0.9996 |
| | 1 | 1e-20 | 0.0374 | 0.9998 | 12.74 | 0.9996 | 9.78e-21 | 0.9996 |
| | 10 | 1e-19 | 0.0374 | 0.9998 | 12.74 | 0.9996 | 9.78e-21 | 0.9996 |
| | 10 | 1e-21 | 0.0374 | 0.9998 | 12.74 | 0.9996 | 9.78e-21 | 0.9996 |
| | 100 | 1e-19 | 0.0374 | 0.9998 | 12.74 | 0.9996 | 9.78e-21 | 0.9996 |
| | 100 | 1e-21 | 0.0374 | 0.9998 | 12.74 | 0.9996 | 9.78e-21 | 0.9996 |
| | 1 | 1e-19 | 0.0374 | 0.9998 | 12.74 | 0.9996 | 9.78e-21 | 0.9996 |
| | 1 | 1e-21 | 0.0374 | 0.9998 | 12.74 | 0.9996 | 9.78e-21 | 0.9996 |

**Supporting Information S5: Experimental procedure and complementary Seebeck coefficient data.**

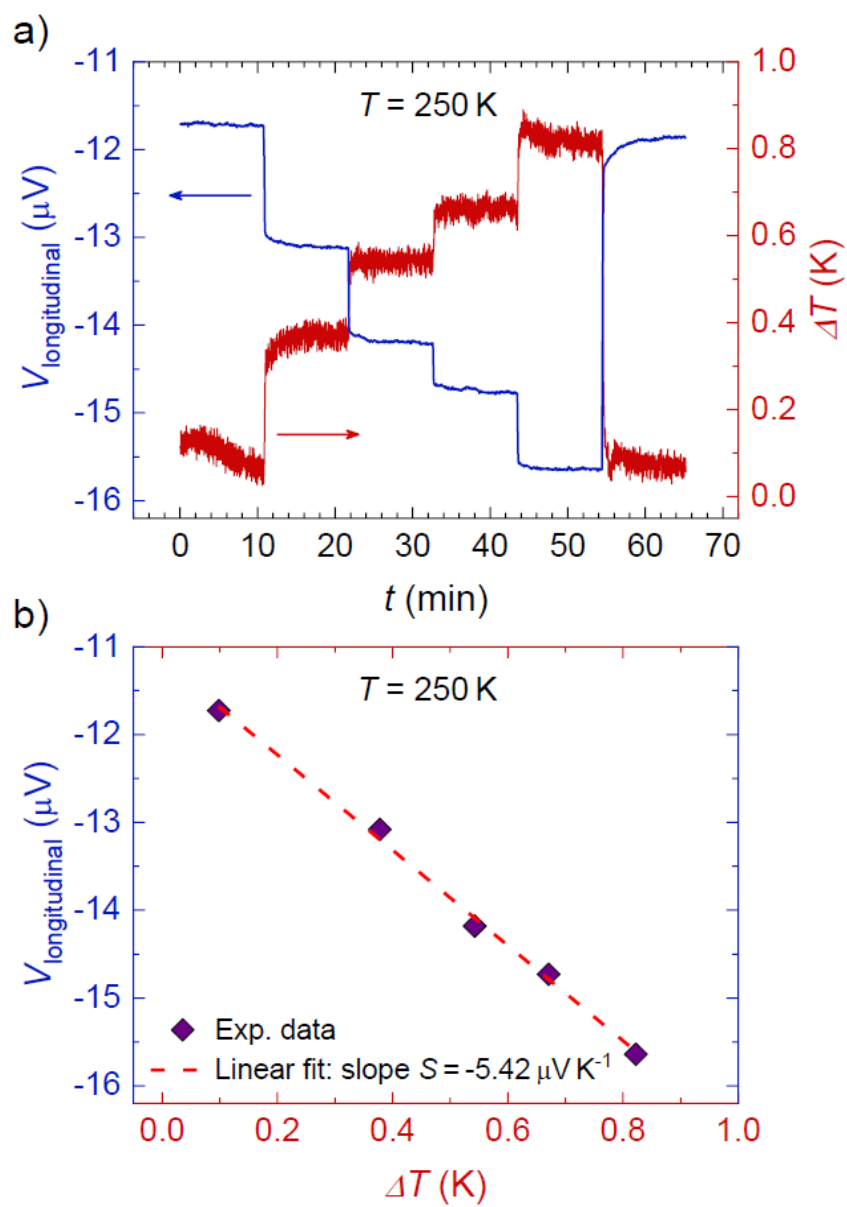

**Figure S5a:** Experimental details of the Seebeck measurements: a) Steps of temperature difference between the Pt resistances and corresponding longitudinal thermoelectric voltages, at a base temperature of 250 K, for one of the SVO films of this study. b) Linear fit of the longitudinal voltage vs temperature difference. The accuracy of the method allows a good measurement of the Seebeck coefficient without increasing much the temperature difference (always lower than 1.5 K), ensuring the reversibility of the process. In this example a Seebeck coefficient of $S = -5.42\ \mu\text{V K}^{-1}$ at $T = 250$ K was extracted. The whole temperature dependence of the Seebeck coefficient for this sample can be seen in Figure S5b (SVO//LSAT, blue curve).

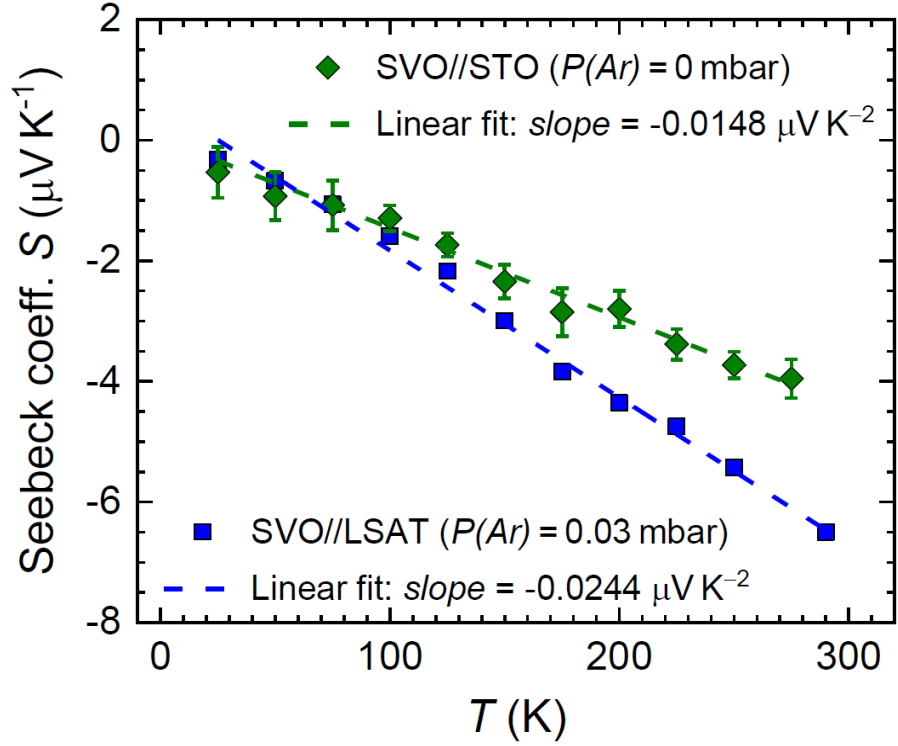

**Figure S5b:** Temperature dependence of the Seebeck coefficient $S$ measured on additional SVO films ($t \approx 70$ nm thick), deposited on LSAT and STO substrates, having different carrier density $n = 2.13 \times 10^{22}$ cm$^{-3}$ and $n = 2.56 \times 10^{22}$ cm$^{-3}$, respectively.

**Supporting Information S6: Softening of selected optical phonon modes with increasing tetragonality in SVO.**

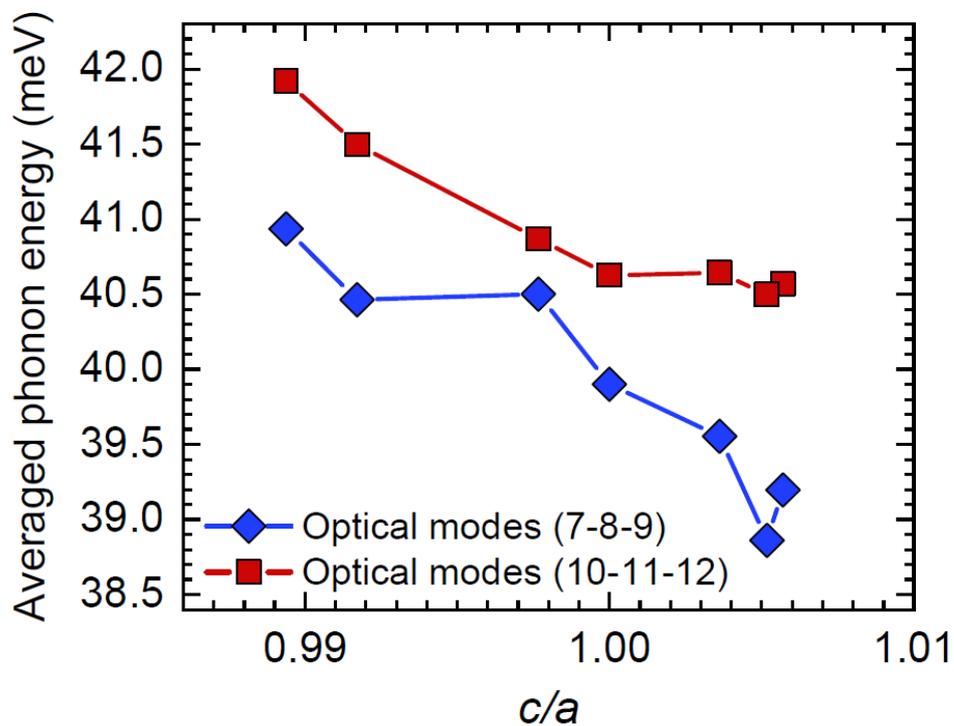

**Figure S6:** Phonons in SVO. Phonon energies at Gamma-point of SVO have been calculated as a function of *c/a* using density functional perturbation theory available in Quantum Espresso. Preliminary data for phonons displaying the strongest e-phonon coupling are shown. The (7-12) indexes refer to optical modes of increasing energy. As the modes 7, 8 and 9, as well as the modes 10, 11 and 12, are split due to the tetragonal distortion, here we plot the averaged values for each group of modes.

**Supporting Information S7: Ellipsometric data.**

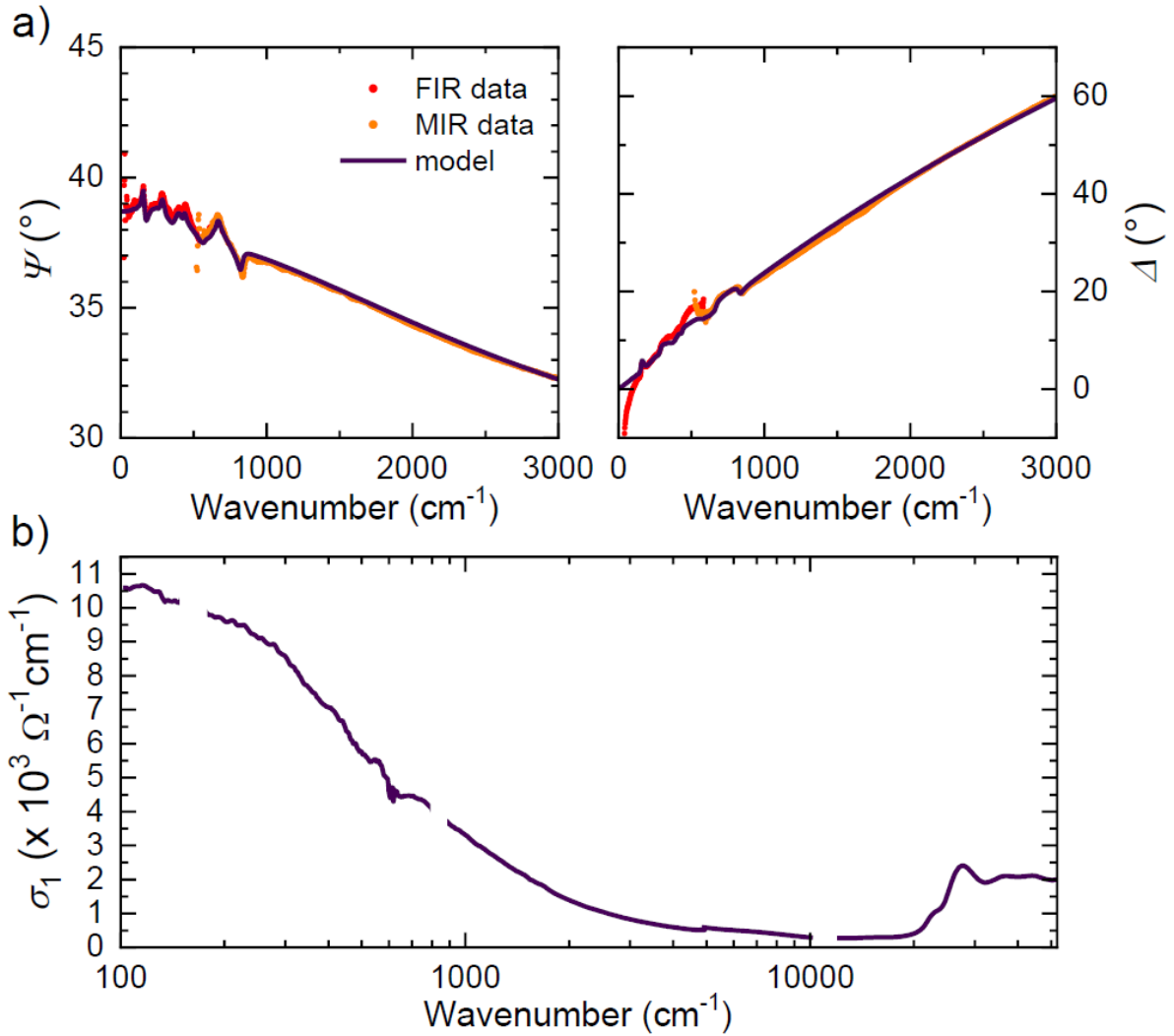

**Figure S7:** a) Ellipsometric data (FIR + MIR) of a SVO film (72 nm thick) deposited on LSAT substrate. The measured ($\Psi$, $\Delta$) spectra are fitted with a substrate/film/ambient model, where only the film response is varied. Substrate response was determined from ellipsometric measurements on a bare substrate.[4] The far-infrared response of the film is dominated by a Drude component. The extracted unscreened plasma frequency is $\omega_p = 19500$ cm$^{-1}$ ($\approx 2.42$ eV) and broadening $\gamma = 680$ cm$^{-1}$, which corresponds to a screened plasma energy $E^*_{\omega_p} = 1.21$ eV (considering $\varepsilon_\infty = 4$, as reported by Makino et al.[5]). The corresponding effective mass $m^* \approx 4.1$ $m_e$. Notice that we reported similar values in our previous studies,[1,6] and that similar values were encountered in literature.[2] Features in the $\Psi/\Delta$ spectra below 1000 cm$^{-1}$ originate from the substrate phonons. b) Optical conductivity $\sigma_1$ of the SVO film up to UV (6.2 eV, 50000 cm$^{-1}$) resulting from point-by-point fit to ellipsometry data.

**Supporting Information S8: Extended universal scaling between the prefactor of the $T^2$ dependent resistivity and the Fermi energy.**

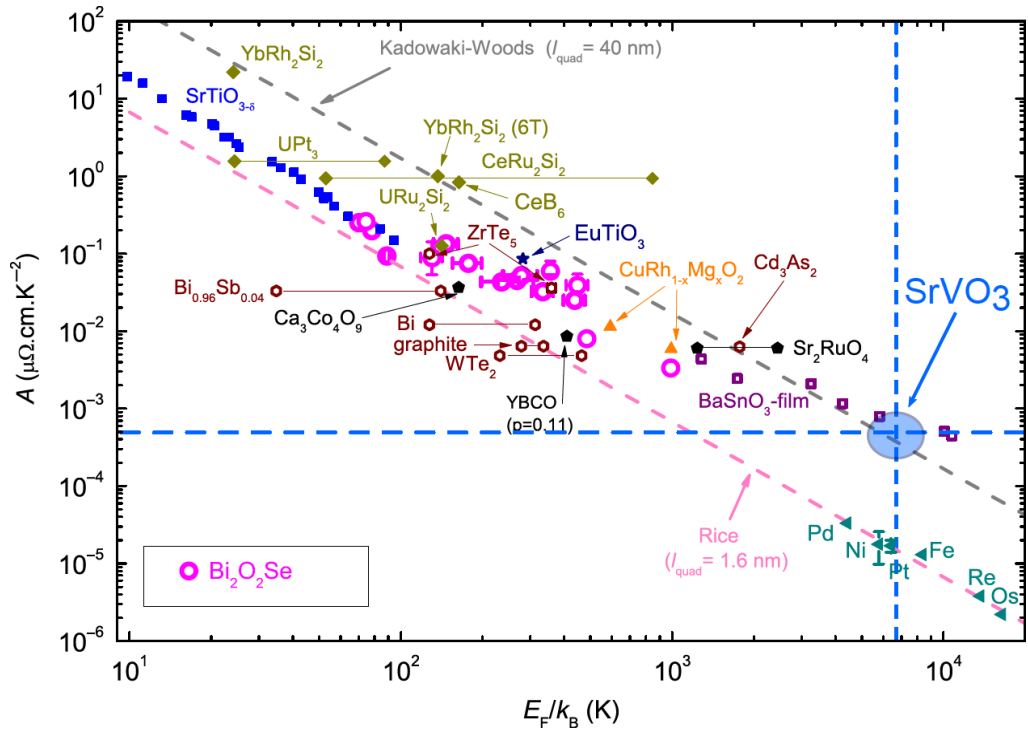

**Figure S8:** Scaling between the prefactor of the $T^2$ dependent resistivity and the Fermi energy. Our data for SVO are plotted on top of data taken from Wang et al.[7]